\UseRawInputEncoding
\documentclass[%
reprint,
superscriptaddress,
amsmath,amssymb,
aps,
]{revtex4-2}

\usepackage{graphicx}
\usepackage{dcolumn}
\usepackage{bm}
\usepackage{soul}
\usepackage[centerlast]{caption}
\usepackage{float}
\usepackage{tabularx}
\usepackage{array}
\usepackage{multirow}
\newcolumntype{Y}{>{\centering\arraybackslash}X}
\usepackage{siunitx}
\usepackage{hyperref}
\hypersetup{colorlinks=true, citecolor=blue, urlcolor=blue, linkcolor=blue}
\usepackage{todonotes}
\usepackage{xcolor}
\usepackage{colortbl}
\makeatletter
\usepackage{braket}
\def\CT@@do@color{%
	\global\let\CT@do@color\relax
	\@tempdima\wd\z@
	\advance\@tempdima\@tempdimb
	\advance\@tempdima\@tempdimc
	\advance\@tempdimb\tabcolsep
	\advance\@tempdimc\tabcolsep
	\advance\@tempdima2\tabcolsep
	\kern-\@tempdimb
	\leaders\vrule
	\hskip\@tempdima\@plus  1fill
	\kern-\@tempdimc
	\hskip-\wd\z@ \@plus -1fill }
\makeatother
\usepackage{colortbl}

\usepackage{soul}
\usepackage{upgreek}

\definecolor{sangria}{rgb}{0.57, 0.0, 0.04}

\begin{document}

\title[Study of Magnetic Field Resilient High Impedance High-Kinetic Inductance Superconducting\\ Resonators]%
{Study of Magnetic Field Resilient High Impedance High-Kinetic Inductance Superconducting Resonators}

\author{C.~Roy}
\affiliation{Institute of Physics, Swiss Federal Institute of Technology (EPFL), 1015 Lausanne, Switzerland.}		
\affiliation{Center for Quantum Science and Engineering, EPFL, 1015 Lausanne, Switzerland.}

\author{S.~Frasca}
\altaffiliation[Also at: ]{Department of Physics, ETH Zurich, CH-8093 Zurich, Switzerland}
\affiliation{Institute of Physics, Swiss Federal Institute of Technology (EPFL), 1015 Lausanne, Switzerland.}		
\affiliation{Center for Quantum Science and Engineering, EPFL, 1015 Lausanne, Switzerland.}

\author{P.~Scarlino}
\email[E-mail: ]{pasquale.scarlino@epfl.ch}
\affiliation{Institute of Physics, Swiss Federal Institute of Technology (EPFL), 1015 Lausanne, Switzerland.}		
\affiliation{Center for Quantum Science and Engineering, EPFL, 1015 Lausanne, Switzerland.}

\date{\today}

\begin{abstract}

Superconducting resonators with high-kinetic inductance play a central role in hybrid quantum circuits, enabling strong coupling with quantum systems with small electric dipole moment and improved parametric amplification. However, optimizing these resonators simultaneously for high internal quality factors ($Q_i$) and resilience to strong magnetic fields remains challenging. In this study, we systematically compare superconducting resonators fabricated from niobium nitride (NbN) and granular aluminum (grAl) thin films, each having similar kinetic inductance values ($L_k \sim 100$~pH/sq). At zero magnetic field, resonators made from grAl exhibit higher $Q_i$ compared to their NbN counterparts. However, under applied magnetic fields, NbN resonators demonstrate significantly better resilience. Moreover, NbN resonators exhibit an unexpected increase in $Q_i$ at intermediate in-plane magnetic fields ($B_{\parallel} \sim 1$~T), which we attribute to an enhanced frequency detuning that reduce coupling to two-level system defects. In contrast, grAl resonators show a distinct critical field above which $Q_i$ rapidly decreases, strongly depending on resonator cross-section respect to the applied field direction. Characterization of the nonlinear properties at zero magnetic field reveals that the self-Kerr coefficient in grAl resonators is more than an order of magnitude higher than in NbN resonators, making grAl particularly attractive for applications requiring pronounced nonlinear interactions. Our findings illustrate a clear trade-off between the two materials: NbN offers superior magnetic-field resilience beneficial for hybrid circuit quantum electrodynamics applications, while grAl is more advantageous in low-field regimes demanding high impedance and strong nonlinearity.
\end{abstract}

\maketitle

\section{Introduction}

Superconducting circuit quantum electrodynamics (cQED) platforms lie at the core of modern quantum technology, owing to their large zero-point fluctuations, low dissipation, strong nonlinearities, and exceptional design versatility~\cite{Blais_2021}. Extensions to the standard cQED architecture have enabled the realization of numerous hybrid quantum systems~\cite{Xiang_2013, Clerk_2020}, in which superconducting microwave cavities coherently interact with mechanical oscillators~\cite{Aspelmeyer_2014}, surface and bulk acoustic wave resonators~\cite{Chu_2018, Satzinger_2018}, magnonic excitations~\cite{Tabuchi_2015}, spin ensembles~\cite{Kubo_2010, Sigillito_2014}, and even individual spins~\cite{Burkard_2023}. These hybrid platforms significantly broaden the scope of systems accessible for quantum computation, sensing, and simulation. However, a significant challenge arises when strong magnetic fields are necessary to define and manipulate quantum degrees of freedom in hybrid cQED devices: the superconducting circuits themselves must preserve coherence under these conditions. Addressing this need for magnetic field-compatible superconducting components is thus the central focus of this work.

A fundamental limitation arises from the fact that conventional superconductors, such as aluminum films, exhibit relatively low critical magnetic fields (on the order of \(10\)~mT~\cite{Al_Bc}), above which superconductivity is suppressed. 
Even type-II superconductors, such as Nb, which can sustain superconductivity up to fields of several tesla~\cite{Al_Bc, Medahinne_2024}, suffer detrimental effects under strong magnetic fields: the superconducting gap decreases, leading to an increased quasiparticle population, and Abrikosov vortices form within the films~\cite{krollMagneticFieldResilient2019, Kwon_2018}. Collectively, these effects introduce microwave resistive losses, significantly compromising device performance. 
Although vortex-trapping techniques, such as flux-pinning holes~\cite{krollMagneticFieldResilient2019, Kwon_2018}, partially mitigate these issues by localizing magnetic flux, they remain insufficient to guarantee robust operation in high magnetic regimes, particularly when the field is oriented perpendicular to the superconducting film.

Additionally, achieving strong coupling in charge-based hybrid systems, particularly for quantum objects with small electric dipole moments~\cite{De_Palma_2024, Stockklauser_2017}, requires superconducting resonators with high characteristic impedance, defined as $Z = \sqrt{\tilde{L}/\tilde{C}}$, where $\tilde{L}$ and $\tilde{C}$ represent the resonator inductance and capacitance per unit length, respectively~\cite{Samkharadze_2016, Stockklauser_2017}.
One possible route to realizing high-impedance resonators is the use of Josephson junction (JJ) arrays~\cite{Masluk_2012}, which provide high Josephson inductance in a compact footprint. However, JJ arrays suffer from intrinsic drawbacks, including strong nonlinearities and high sensitivity to magnetic fields~\cite{Kuzmin_2023}. These limitations hinder their applicability in hybrid quantum devices requiring operation in high magnetic fields.

A promising alternative is the use of high-kinetic inductance disordered superconducting thin films, such as NbTiN~\cite{Yu_2021,Bahr_2024}, TiN~\cite{Joshi_2022, Shearrow_2018}, NbN~\cite{frasca_2023,niepceHighKineticInductance2019}, and granular aluminum (grAl)~\cite{maleeva_2018,grunhauptGranularAluminumSuperconducting2019,borisovSuperconductingGranularAluminum2020}. These materials exhibit a reduced Cooper pair density, resulting in a significant kinetic inductance contribution to the total inductance leading to increased impedance and reduced phase velocity~\cite{Annunziata_2010}. NbTiN, NbN and TiN typically form polycrystalline films, whereas grAl consists of pure aluminum grains embedded within an oxide matrix~\cite{Glezer_Moshe_2020_nbn_vs_gral}.

Beyond their magnetic-field resilience, these materials exhibit intrinsic nonlinearities arising from the quadratic dependence of the kinetic inductance $L_k$ on the current $I$~\cite{Annunziata_2010}. Although this distributed nonlinearity is typically several orders of magnitude smaller than that achievable with Josephson junctions, it can nonetheless be leveraged for various quantum applications, including qubit architectures~\cite{winkel_2020}, parametric amplifiers~\cite{frasca_2024,Zapata_2024}, frequency converters~\cite{Khalifa_2024}, and single-photon detectors~\cite{Ognjen_2024}. Their higher critical magnetic fields $B_C$, increased critical temperatures $T_C$, and controllable high inductance make high-kinetic inductance thin films promising candidates for a wide spectrum of superconducting quantum technologies.

In this study, we investigate the internal quality factor, nonlinear response, and magnetic-field resilience of superconducting resonators fabricated from two high-kinetic-inductance thin films: NbN and grAl. The internal quality factor ($Q_i$) of a resonator is a crucial parameter to assess coherence in quantum devices, while the resonator's intrinsic nonlinearity plays a central role in defining qubits, mediating photon-photon interactions and supporting parametric amplification~\cite{wallraff_2004,zimmer_1967,Yurke_2006,Kochetov_2015,Peacock_1996}. By evaluating these properties under strong magnetic fields, we provide insights on the suitability and trade-offs associated with NbN and grAl for future hybrid quantum devices and advanced superconducting circuit applications.

\section{Main parameters, challenges, and goals}

\begin{figure}[h!]
\centering
\includegraphics[width=\linewidth]{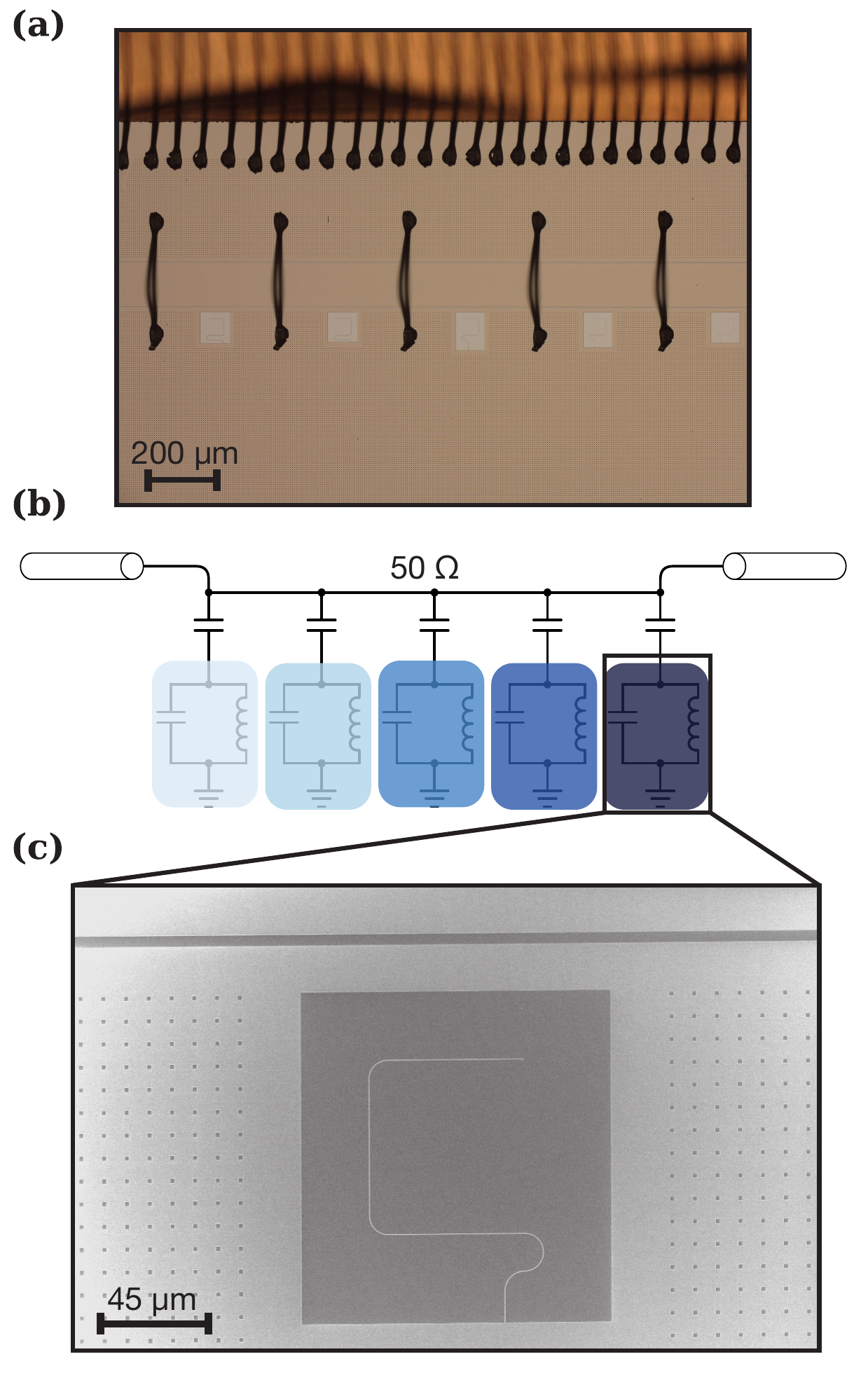}
\caption{\textbf{Representative device.} 
\textbf{(a)} Optical microscope image of the measured grAl device, showing the $50~\Omega$ signal feedline and the etched windows hosting five suspended nanowire resonators. 
\textbf{(b)} Circuit schematic corresponding to the portion of the device shown in \textbf{(a)}, with the nanowire resonators modeled as LC circuits shunted to ground. Resonator widths range from 600~nm (left) to 200~nm (right) and are indicated by a blue gradient. 
\textbf{(c)} Scanning electron microscope (SEM) image of a 200~nm wide grAl resonator also showing the etched flux traps in the ground plane designed to mitigate vortex formation.}
\label{fig:device}
\end{figure}

We present the characterization of two superconducting films with similar kinetic inductance ($L_k$): a 13~nm NbN film with $L_k = 89$~pH/sq and a 50~nm granular aluminum (grAl) film with $L_k = 150$~pH/sq. While ideally, a direct comparison would involve films with identical $L_k$, the inherent variability in the fabrication of sputtered grAl films presents a challenge. Consequently, we selected the grAl film with the closest $L_k$ to our NbN film. This approach allows us to probe the high-end of achievable $L_k$ values for NbN and the lower-end values characteristic of grAl, enabling an effective evaluation of resonator quality factors at similar characteristic impedances.

\begin{figure*}
\centering
\includegraphics[width=.96\linewidth]{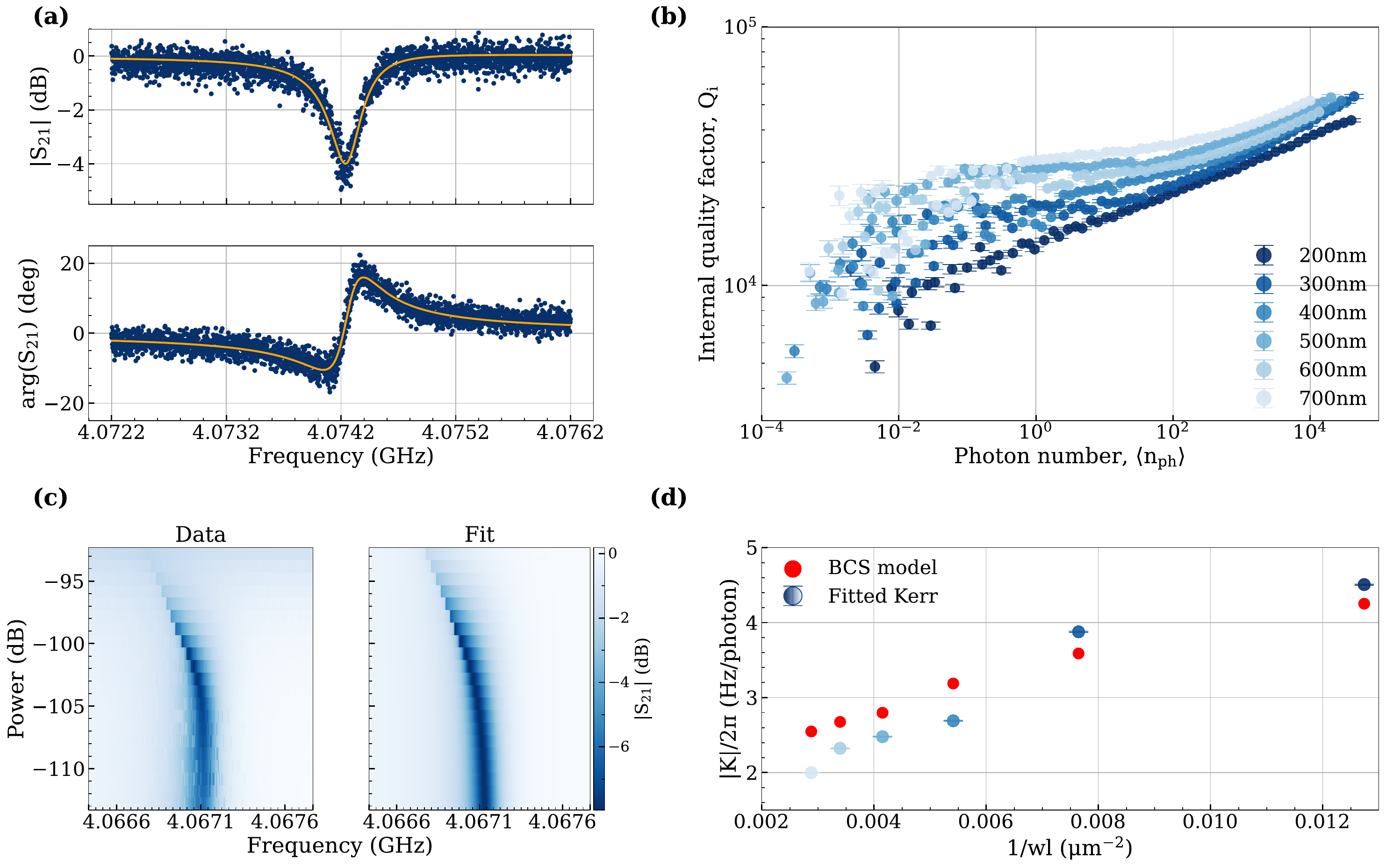}
\caption{\textbf{Characterization of NbN resonators.} 
\textbf{(a)} Representative spectroscopy. Amplitude ($|S_{21}|$) and phase ($\arg(S_{21})$) response of the microwave feedline for the 200~nm wide resonator, recorded at an average intracavity photon number $\langle n_{ph} \rangle \sim 1$. The orange line represents the fit to Eq.~\eqref{eq:S21}, performed directly on the complex data (see Appendix~\ref{app:spectroscopy_fit}). 
\textbf{(b)} Power dependence of the internal quality factor. Internal quality factors ($Q_i$) as a function of $\langle n_{ph} \rangle$ for all NbN resonators at zero magnetic field are extracted according to Eq.~\eqref{eq:S21} and extend up to the onset of nonlinearity ($K \langle n_{ph} \rangle \approx \bar{\kappa}$, where $\bar{\kappa}$ is the total loss rate of the resonator).
\textbf{(c)} Nonlinearity study using a 2D power map. Measured 2D power sweep of the 200~nm wide resonator, showing the variation of $|S_{21}|$ and resonator resonant frequency with increasing power delivered to the device. The self-Kerr coefficient ($K$) is extracted from the fit using Eq.~\eqref{eq:S21_nonlin}. The onset of the Duffing effect is visible as an effective resonant frequency redshift. 
\textbf{(d)} Self-Kerr coefficient ($K$). Comparison of the extracted self-Kerr coefficient $K$ (blue gradient) with theoretical estimates from the BCS model (red points) using Eq.~\eqref{eq:Kerr_BCS}. Error bars represent fitting uncertainties.}
\label{fig:nbn_characteristics}
\end{figure*}

The resonators are designed to maximize resilience to out-of-plane magnetic fields. Each device consists of five overcoupled ($\kappa > \gamma$, where $\kappa$ and $\gamma$ are the resonator external and internal losses, respectively) $\lambda/4$ wire resonators fabricated in a hanged configuration, as illustrated in Fig.~\ref{fig:device}\textbf{(a-b)}. The overcoupled nature of the resonators ensures that their resonances remain trackable and measurable during the magnetic field sweep, where a reduction in the resonator internal quality factor ($Q_i$) is expected. These resonators are capacitively coupled to a common 50~$\Omega$ feedline, with resonant frequencies spanning 4 to 7~GHz and wire widths $w$ ranging from 700~nm to 200~nm in a 100~nm step. The corresponding resonator impedances range from 1.5~k$\Omega$ to 3.5k~$\Omega$ (see Tab.~\ref{table:NbN_resonators} and Tab.~\ref{table:grAl_resonators}). A representative SEM image of a grAl resonator is shown in Fig.~\ref{fig:device}\textbf{(c)}. Due to the different $L_k$ values of the two films, the resonator lengths $l$ are adjusted to achieve comparable resonant frequencies for a given wire width $w$. To mitigate the formation of magnetic-field-induced vortices, flux traps are incorporated into the ground plane~\cite{krollMagneticFieldResilient2019}. Further details on the experimental setup and packaging are provided in Appendix~\ref{app:setup}.

We investigate the $Q_i$ of the resonators both in the absence and presence of an applied magnetic field. The two dominant loss mechanisms in superconducting resonators are described by the two-level system (TLS) model and quasiparticle-induced dissipation~\cite{scigliuzzo_2020,Pappas_2011}:
\begin{equation}
    \frac{1}{Q_i} = F \delta_{TLS}^{0} \frac{\tanh[\hbar \omega_0 /(2k_B T)]}{(1+\langle n_{ph} \rangle /n_C)^\beta} + \frac{\alpha}{\pi} \sqrt{\frac{2\Delta}{hf_r}} \frac{n_{qp}(T)}{n_s(0)\Delta} + \delta_0.
    \label{eq:QTLS}
\end{equation}
In the TLS model, $F$ represents the filling factor (i.e., the ratio between the electric field interacting with TLSs and the total electric field), $\delta_{TLS}^{0}$ is the intrinsic TLS loss, and $n_C$ is the characteristic critical intracavity photon number at which TLS saturation occurs. For quasiparticle-induced losses, $\alpha$ represents the ratio of kinetic inductance $L_k$ to total inductance $L_t$ of the resonator. In these devices, the kinetic term dominates, so the geometric contribution is negligible and $\alpha \approx 1$. The term $n_{qp}(T)$ represents the temperature-dependent quasiparticle population, while $n_s(0)$ is the zero-energy density of states of Cooper pairs. Finally, $\delta_0$ accounts for residual losses not captured by these TLS and quasiparticle models.

\begin{figure*}
\centering
\includegraphics[width=.96\linewidth]{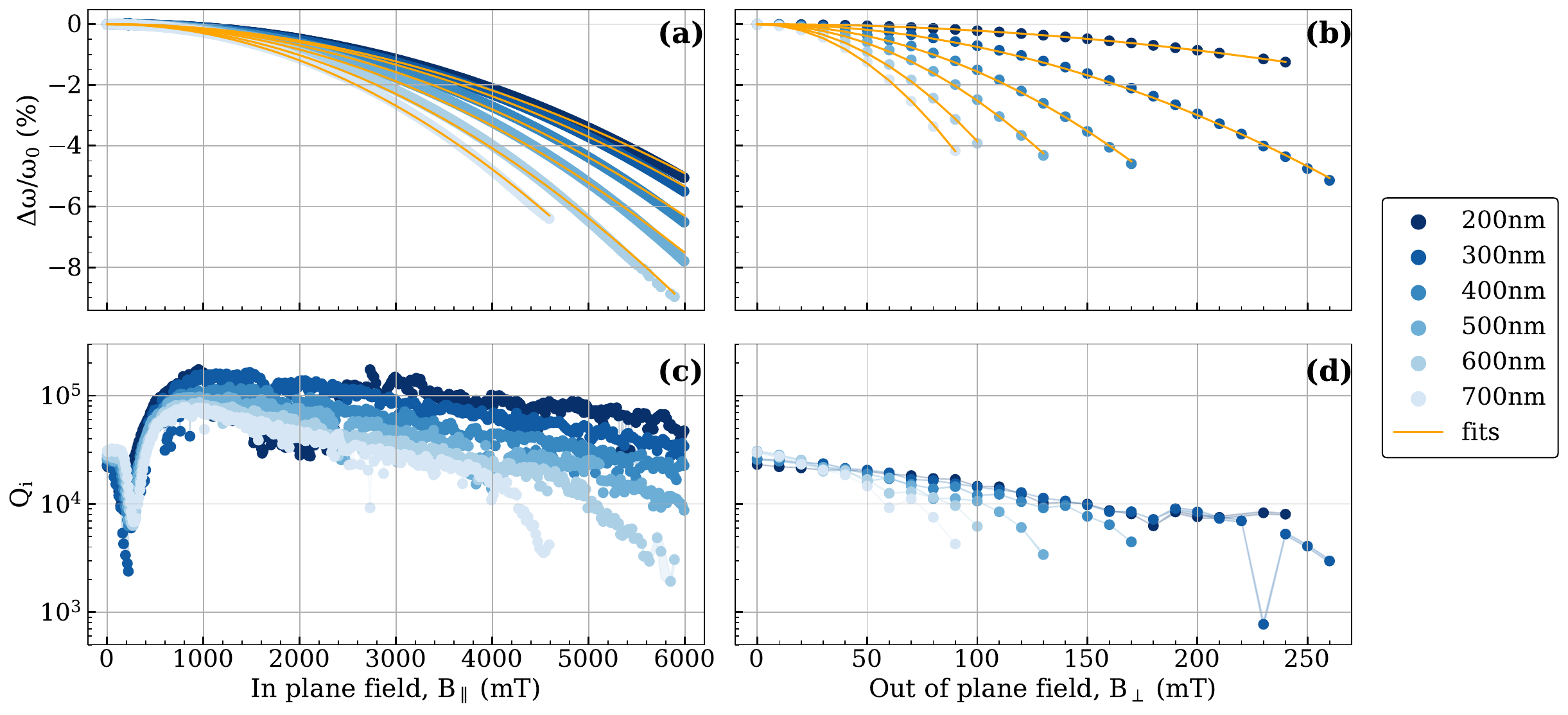}
\caption{\textbf{Magnetic field response of NbN resonators.}  
\textbf{(a)} Normalized resonant frequency shift as a function of in-plane magnetic field. The plot shows the relative frequency shift $\Delta \omega / \omega_0$ for all NbN resonators under an applied in-plane field $B_\parallel$. The fitted curves (orange lines), based on Eq.~\eqref{eq:BC}, allow extraction of the resonators' critical in-plane magnetic fields $B_{C,\parallel}$. The observed deviations between different resonators are attributed to a magnet misalignment of approximately $1^\circ$ out-of-plane. The field sweep extends up to 6~T, the maximum available field of the vector magnet.  
\textbf{(b)} Normalized resonant frequency shift as a function of out-of-plane magnetic field. The extracted resonator critical out-of-plane magnetic fields $B_{C,\perp}$ are determined from the fits (orange lines) based on Eq.~\eqref{eq:BC}.  
\textbf{(c)} Evolution of the internal quality factor $Q_i$ with increasing in-plane magnetic field $B_\parallel$ for all six NbN resonators.  
\textbf{(d)} Evolution of the internal quality factor $Q_i$ with increasing out-of-plane magnetic field $B_\perp$ for all six NbN resonators.}
\label{fig:nbn_bfield}
\end{figure*}

The nonlinearity of the devices is characterized in the absence of a magnetic field. The kinetic inductance dependence on current, in the limit of $T/T_C < 0.1$, follows the relation~\cite{Clem_2012}:
\begin{equation}
    L_k(I) \approx L_k(0) \left[ 1 - \left( \frac{I}{I_*} \right)^n \right]^{-1/n},
    \label{eq:Lk_withI}
\end{equation}
where $I_*$ is the critical depairing current and $n=2.21$.
This dependence leads to a distributed nonlinearity across the entire film, which can be quantified for resonators by the self-Kerr term $K$. Within the Bardeen–Cooper–Schrieffer (BCS) theory for disordered superconductors, $K$ is given by~\cite{anferov_2020,eichler_2014}:
\begin{equation}
    K = -\frac{3}{8} \frac{\hbar \omega_r^2}{L_t I_*^2},
    \label{eq:Kerr_BCS}
\end{equation}
where $\omega_r$ is the resonator resonant angular frequency, and $L_t$ is the total inductance, which is related to the kinetic inductance via $L_t = L_k (l/w)$, with $l$ and $w$ denoting the resonator length and width, respectively. The self-Kerr nonlinearity $K$ is expected to scale linearly with the inverse of the length-width product, $1/(wl)$, as derived in Appendix~\ref{app:selfK-est}. To accurately estimate $K$, precise values of both the resonator resonant frequency (extracted from measurements) and the total resonator inductance $L_t$ are required. The latter is verified using numerical simulations in Sonnet~\footnote{https://www.sonnetsoftware.com/} and Ansys~\footnote{https://www.ansys.com/}. However, determining the critical current $I_*$ of the resonator wires remains challenging~\cite{frasca_2019}. In this work, the critical current is extracted following the method described in Ref.~\cite{frasca_2024}. 

Studying the resonator critical field $B_C$ and the evolution of the internal quality factor $Q_i$ under an applied magnetic field is essential for optimizing high-$L_k$ films in hybrid quantum devices. When a magnetic field is applied, vortices can nucleate within the superconducting film, acting as dissipation channels through enhanced quasiparticle generation~\cite{krollMagneticFieldResilient2019}, leading to performance degradation. The characteristic size of these vortices is on the order of the London penetration depth, $\lambda_p$~\cite{Sonier_2004}. In thin films, where the thickness is smaller than the penetration depth ($t < \lambda_p$), vortex formation is significantly suppressed, improving the magnetic field resilience of the devices~\cite{krollMagneticFieldResilient2019}. The London penetration depth for NbN is estimated to be approximately 200~nm~\cite{Khan_2022}, whereas for grAl, it is closer to 500~nm~\cite{Abraham_1978}. 

Despite the suppression of vortices, the superconducting energy gap $\Delta(B)$ decreases with increasing magnetic field, following the relation: $\Delta(B) = \Delta_0 \sqrt{1 - (B/B_C)^2}$~\cite{tinkham2004introduction}. A reduced superconducting gap increases the susceptibility of Cooper pairs to breaking, leading to a lower Cooper pair density and a corresponding increase in the quasiparticle population. To maintain a supercurrent under these conditions, the remaining Cooper pairs must carry additional kinetic energy, resulting in an increased $L_k$ and a quadratic reduction in the resonator frequency~\cite{Clem_2012}. For out-of-plane magnetic fields, vortex nucleation occurs within the resonator (typically when $\lambda_p < w$), in the feedline and in the ground plane. Although flux traps are implemented to mitigate vortex motion in the ground plane~\cite{krollMagneticFieldResilient2019, Kwon_2018}, they are generally insufficient to eliminate their effects entirely. Consequently, thin-film devices tend to exhibit higher resilience to in-plane magnetic fields than to out-of-plane fields. A key parameter for assessing and comparing magnetic field resilience across different device geometries and material platforms is the resonator critical field, $B_C$.
    
The normalized shift in the resonant frequency of the resonator, $\Delta \omega / \omega_0$ (where $\omega_0$ is the resonator resonant frequency at zero magnetic field, $B_{\parallel,\perp}=0$), induced by an in-plane magnetic field $B_\parallel$, can be modeled using the following relation~\cite{Samkharadze_2016}:
\begin{equation}
    \Delta \omega / \omega_0 = -\frac{\pi}{48} \frac{e^2t^2}{\hbar k_B T_C} D \bigg( 1+\theta_B^2 \frac{w^2}{t^2} \bigg) B_{\parallel}^2,
    \label{eq:freq_shift_bfield_inplane}
\end{equation}
where $t$ is the film thickness, $T_C$ is the critical temperature, $w$ is the resonator width, and $\theta_B$ represents the out-of-plane misalignment angle of the applied field. The presence of misalignment introduces a $w$ dependence to $\Delta \omega / \omega_0$, which would otherwise depend solely on the film thickness $t$. Following Ref.~\cite{Samkharadze_2016}, the frequency shift can be approximated as:
\begin{equation}
    \Delta \omega / \omega_0 \approx - \frac{1}{4} \bigg( \frac{B}{B_C} \bigg)^2,
    \label{eq:BC}
\end{equation}
from which $B_C$ can be extracted.

Throughout this work, resonators are color-coded for clarity: NbN resonators of varying widths are represented using a blue gradient, while grAl resonators are shown using a green gradient.

\section{NbN devices}

\subsection{\label{sec:intro}Device characterization at optimal conditions}

The sputtered NbN film used in this study is 13~nm thick, with a kinetic inductance of $L_k=89$~pH/sq. This relatively high $L_k$ value was chosen to closely match that of the grAl film, enabling a more meaningful comparison between the two materials. Details of the film fabrication, which follows established protocols~\cite{dane_bias_2017, frasca_2023}, and the $L_k$ estimation procedure using Sonnet, can be found in Appendix~\ref{app:Lk_est}.

The NbN resonators were first characterized under optimal conditions (i.e., at zero magnetic field) to establish baseline performance metrics. A representative resonance spectrum is shown in Fig.~\ref{fig:nbn_characteristics}\textbf{(a)}, depicting the transmission scattering parameter $S_{21}$ of a 200~nm wide resonator coupled to the feedline at an estimated input power corresponding to an average intracavity photon number $\langle n_{ph} \rangle \sim 1$, along with its corresponding fit. Details of the fitting procedure are provided in Appendix~\ref{app:spectroscopy_fit}. To further investigate the evolution of $Q_i$ as a function of $\langle n_{ph} \rangle$ and to analyze dominant loss mechanisms, a scan of the power delivered to the device was conducted for each resonator, as shown in Fig.~\ref{fig:nbn_characteristics}\textbf{(b)}. For simplicity, in the rest of the manuscript, we will refer to similar measurements as "power scan".

At $\langle n_{ph} \rangle < 1$, the $Q_i$ of the resonators is approximately $1 \times 10^4$, consistent with previous studies~\cite{frasca_2023}. As $\langle n_{ph} \rangle$ increases, the saturation of TLS losses leads to an increase in $Q_i$, reaching approximately $5 \times 10^4$. However, the expected high-power saturation of $Q_i$ is not fully reached due to the onset of Duffing nonlinearity~\cite{anferov_2020}, which is not included in this fit. In particular, a correlation is observed between resonator inductor width $w$ and $Q_i$, with wider wires exhibiting higher $Q_i$ values on average. For measurements conducted below an $\langle n_{ph} \rangle \sim 1$, the reliability of extracted $Q_i$ values is limited due to the low signal-to-noise ratio.

To characterize the resonator’s nonlinearity, we perform a global fit to the two-dimensional complex $S_{21}$ spectroscopy map, measured as a function of probe frequency and input power delivered to the device [see Fig.~\ref{fig:nbn_characteristics}\textbf{(c)}]. Following the procedure described in Appendix~\ref{app:selfK-est}, this fitting allows extraction of the resonator’s self-Kerr coefficient, $K$. In the high-photon regime, where nonlinear effects become non-negligible ($K \langle n_{ph} \rangle \approx \bar{\kappa}$, where $\bar{\kappa} = \kappa + \gamma$), the resonator resonant frequency shift becomes more pronounced. The experimentally extracted $K$ values are compared in Fig.~\ref{fig:nbn_characteristics}\textbf{(d)} to those predicted by Eq.~\eqref{eq:Kerr_BCS}. The critical currents $I_*$ required for this estimation are extracted from a reference device fabricated on the same film following the method from Ref.~\cite{frasca_2024}. These values range from $I_* = 260$~$\mu A$ for the 700~nm wide resonator to $I_* = 74$~$\mu A$ for the 200~nm wide resonator. The results confirm that the fitted $K$ values follow the expected linear dependence on the inverse of the length-width product ($1/lw$) of the resonator, as predicted by Eq.~\eqref{eq:Kerr_BCS}, with narrower resonators exhibiting stronger nonlinearity. Discrepancies between the fitted data and the theoretical predictions based on the BCS model likely arise from uncertainties in input power estimation or variations in critical current, but these deviations remain limited. The theoretical model produces values within the correct order of magnitude, validating the use of these estimated parameters for future device designs.

\subsection{\label{sec:nbn_bfield_chara}Characterization in high magnetic field}

\begin{figure*}
\centering
\includegraphics[width=.96\linewidth]{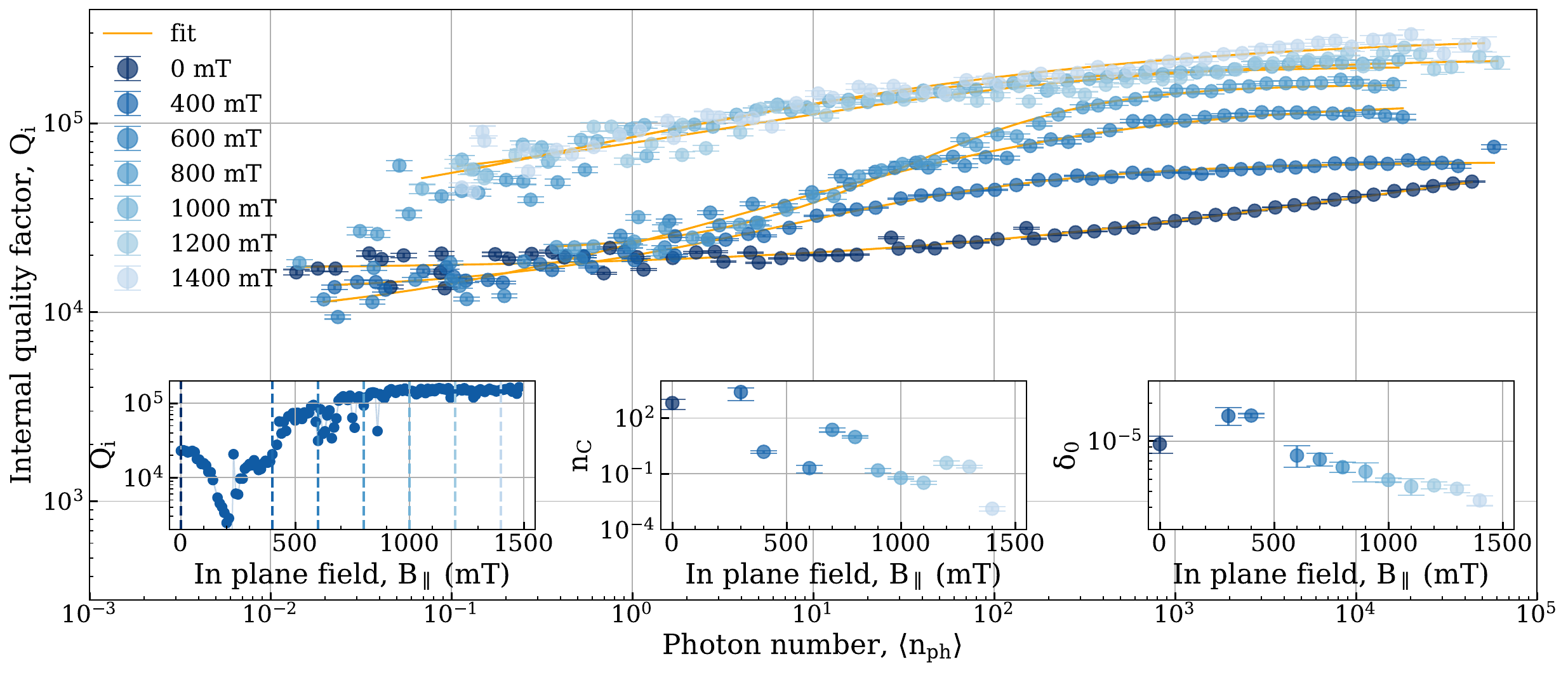}
\caption{\textbf{Power scans of a NbN resonator at different in-plane magnetic fields.} Plot of the extracted internal quality factor $Q_i$ of the 300~nm wide NbN wire resonator (according to Eq.~\eqref{eq:S21}) as a function of the average intracavity photon number ($\langle n_{ph} \rangle$) for different in-plane magnetic field values ($B_\parallel$). The left inset, previously reported in Fig.~\ref{fig:nbn_bfield}\textbf{(c)}, represents the evolution of extracted $Q_i$ with respect to $B_\parallel$. The vertical dashed lines indicates the $B_\parallel$ values at which the power scans were performed. The fitted curves (orange lines), based on Eq.~\eqref{eq:QTLS}, are shown alongside the data. The middle inset displays the fitted critical intracavity photon number ($n_C$) extracted from each power scan. The right inset presents the fitted loss rate after TLS saturation ($\delta_0$) as a function of $B_{\parallel}$.}
\label{fig:nbn_pscan_bfield}
\end{figure*}

\begin{figure*}
\centering
\includegraphics[width=.96\linewidth]{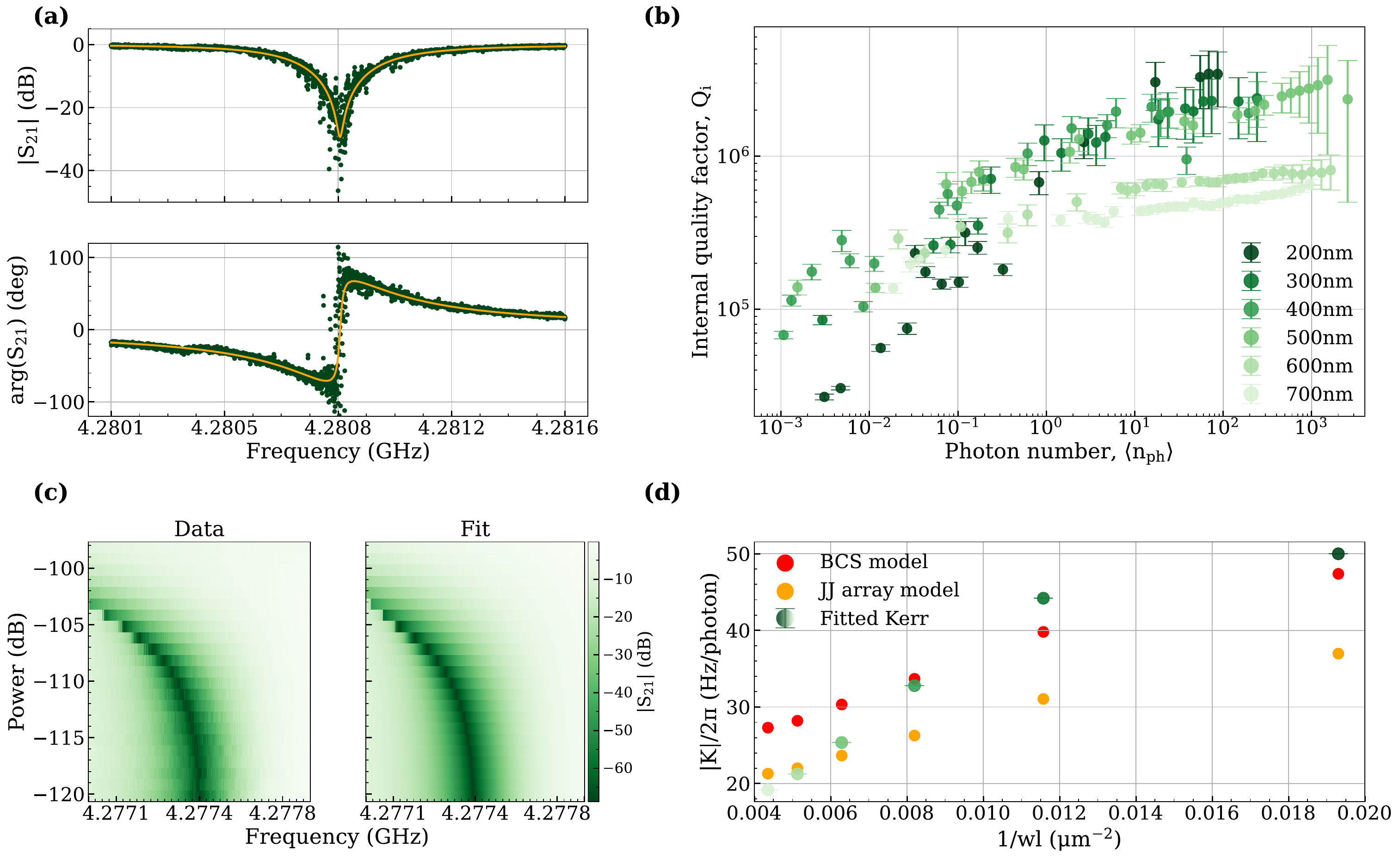}
\caption{\textbf{Characterization of grAl resonators.} 
\textbf{(a)} Representative spectroscopy. Amplitude ($|S_{21}|$) and phase ($\arg(S_{21})$) response of the microwave feedline for the 200~nm wide resonator, recorded at an average intracavity photon number $\langle n_{ph} \rangle \sim 1$. The orange line represents the fit to Eq.~\eqref{eq:S21}, performed directly on the complex data (see Appendix~\ref{app:spectroscopy_fit}). 
\textbf{(b)} Power dependence of the internal quality factor. Internal quality factors ($Q_i$) as a function of $\langle n_{ph} \rangle$ for all NbN resonators at zero magnetic field are extracted according to Eq.~\eqref{eq:S21} and extend up to the onset of nonlinearity ($K \langle n_{ph} \rangle \approx \bar{\kappa}$, where $\bar{\kappa}$ is the total loss rate of the resonator).
\textbf{(c)} Nonlinearity study using a 2D power map. Measured 2D power sweep of the 200~nm wide resonator, showing the variation of $|S_{21}|$ and resonator resonant frequency with increasing power delivered to the device. The self-Kerr coefficient ($K$) is extracted from the fit using Eq.~\eqref{eq:S21_nonlin}. The onset of the Duffing effect is visible as an effective resonant frequency redshift. 
\textbf{(d)} Self-Kerr coefficient ($K$). Comparison of the extracted self-Kerr coefficient $K$ (green gradient) with theoretical estimates from the BCS model (red points) using Eq.~\eqref{eq:Kerr_BCS} and from the JJ array model (orange points) using Eq.~\eqref{eq:Ker_JJ}. Error bars represent fitting uncertainties.}
\label{fig:gral_characteristics}
\end{figure*}

The resonator’s critical fields $B_C$ are determined from the normalized resonant frequency shift \(\Delta \omega / \omega_0\) as a function of the applied fields $B_\parallel$ and $B_\perp$, shown in Fig.~\ref{fig:nbn_bfield}\textbf{(a-b)}, and fitted to Eq.~\eqref{eq:BC}. These measurements were conducted with an input power corresponding to an average intracavity photon number of $\langle n_{ph} \rangle \sim 100$ (at $B=0$). Assuming that the effective cross-sectional area exposed to the in-plane magnetic field $B_\parallel$ is similar for all resonators, the normalized resonant frequency shifts $\Delta \omega / \omega_0$ are expected to overlap for all the six studied resonators. The deviation observed in Fig.~\ref{fig:nbn_bfield}\textbf{(a)} is attributed to a small out-of-plane magnetic field component, with an estimated misalignment angle of $\theta_B = 1.03^\circ$ (see Appendix~\ref{app:bfield_misalignment} for details). 
Due to the overcoupled nature of the resonances at this $\langle n_{ph} \rangle$, we can track the resulting frequency shifts up to 6~T, corresponding to the maximum field of our vector magnet.
The extracted in-plane resonator critical fields range from $B_{C,\parallel} = (9.148 \pm 0.006)$~T for the 700~nm wide resonator to $B_{C,\parallel} = (13.537 \pm 0.009)$~T for the 200~nm wide resonator (additional details in Table~\ref{table:NbN_resonators}).

For out-of-plane magnetic fields $B_\perp$, $\Delta \omega / \omega_0$ exhibits a strong dependence on the resonator width $w$, as shown in Fig.~\ref{fig:nbn_bfield}\textbf{(b)}. The extracted out-of-plane critical fields indicate a maximum resilience of $B_{C,\perp} = (1076.6 \pm 0.6)$~mT for the 200~nm wide resonator, whereas the 700~nm wide resonator exhibits a significantly lower $B_{C,\perp} = (220.1 \pm 0.6)$~mT.

While $B_C$ provides an estimate of a resonator's resilience to applied magnetic fields, it does not account for the effects of losses. To address this aspect, we monitor $Q_i$ as a function of both in-plane and out-of-plane magnetic fields. For the in-plane case [see Fig.~\ref{fig:nbn_bfield}\textbf{(c)}], the evolution of $Q_i$ can be divided into three distinct regions. Initially, a sharp dip in $Q_i$ occurs around 0.3~T, with the precise field value depending on the resonator’s resonant frequency. This feature is attributed to coupling with paramagnetic spin impurities in the silicon substrate, leading to an electron spin resonance (ESR) dip~\cite{Samkharadze_2016,Yu_2021,Bahr_2024}. Following this ESR-induced dip, $Q_i$ exhibits a sharp increase, surpassing its zero-field value and reaching or even overcoming $Q_i \sim 1 \times 10^5$. Beyond 1~T, $Q_i$ exponentially decreases as the field approaches the critical value. Notably, at 6~T, the 200~nm wide resonator retains a relatively high quality factor of $Q_i \sim 4 \times 10^4$, demonstrating the robustness of nanowire high-$L_k$ resonators under strong magnetic fields.
Throughout the field sweep, sharp dips in $Q_i$ are observed at various field values, but recover over time, likely due to vortex-induced losses. Additionally, the 200~nm wide resonator exhibits a broader decrease in $Q_i$ between 1.5~T and 2.5~T, likely caused by coupling to standing waves in the feedline. The evolution of the coupling quality factor $Q_c$ can be found in Appendix~\ref{app:bfield_qc}.

To gain further insight into the improvement in $Q_i$ following the ESR-induced dip, we perform detailed power scans on the 300~nm wide resonator at specific field values near this region [see Fig.~\ref{fig:nbn_pscan_bfield}]. For clarity, not all power scans are displayed. For each studied $B_{\parallel}$ value, a resonator power scan is fitted using Eq.~\eqref{eq:QTLS}, and key fitting parameters, including $n_C$ and $\delta_0$, are extracted. The results reveal a significant decrease in $n_C$ by several orders of magnitude as the in-plane field increases. Additionally, the residual loss rate $\delta_0$ after TLS saturation is reduced by nearly an order of magnitude compared to the zero-field case. This suggests that increasing the magnetic field may suppress TLS effects by detuning magnetic-dependent TLS far from the resonator frequencies, thereby reducing the density of TLS to saturate and increasing $Q_i$ for a given $\langle n_{ph} \rangle$~\cite{Bahr_2024, Zollitsch_2019}. Another possible explanation relies on the presence of Abrikosov vortices in the superconducting film, which could act as quasiparticle traps, thereby mitigating losses due to quasiparticle generation~\cite{krollMagneticFieldResilient2019}. However, this hypothesis appears unlikely in our case, as the applied magnetic field is in-plane—where vortex formation is suppressed—and flux traps are implemented to mitigate effects from the small out-of-plane component caused by magnet misalignment. The observed increase of $Q_i$ at finite magnetic fields could be leveraged to mitigate resonator losses by saturating TLS at lower input power, potentially improving the performance of superconducting devices.

For out-of-plane magnetic fields, $Q_i$ degradation is significantly more pronounced and strongly dependent on resonator width $w$ [see Fig.~\ref{fig:nbn_bfield}\textbf{(d)}]. For instance, the 200~nm wide resonator reaches an internal quality factor of $Q_i \sim 8 \times 10^3$ before becoming too undercoupled to be observed beyond 250~mT. This result highlights the expected reduced resilience of resonators to out-of-plane fields compared to in-plane fields.
   
\section{grAl device}

\begin{figure*}
\centering
\includegraphics[width=.96\linewidth]{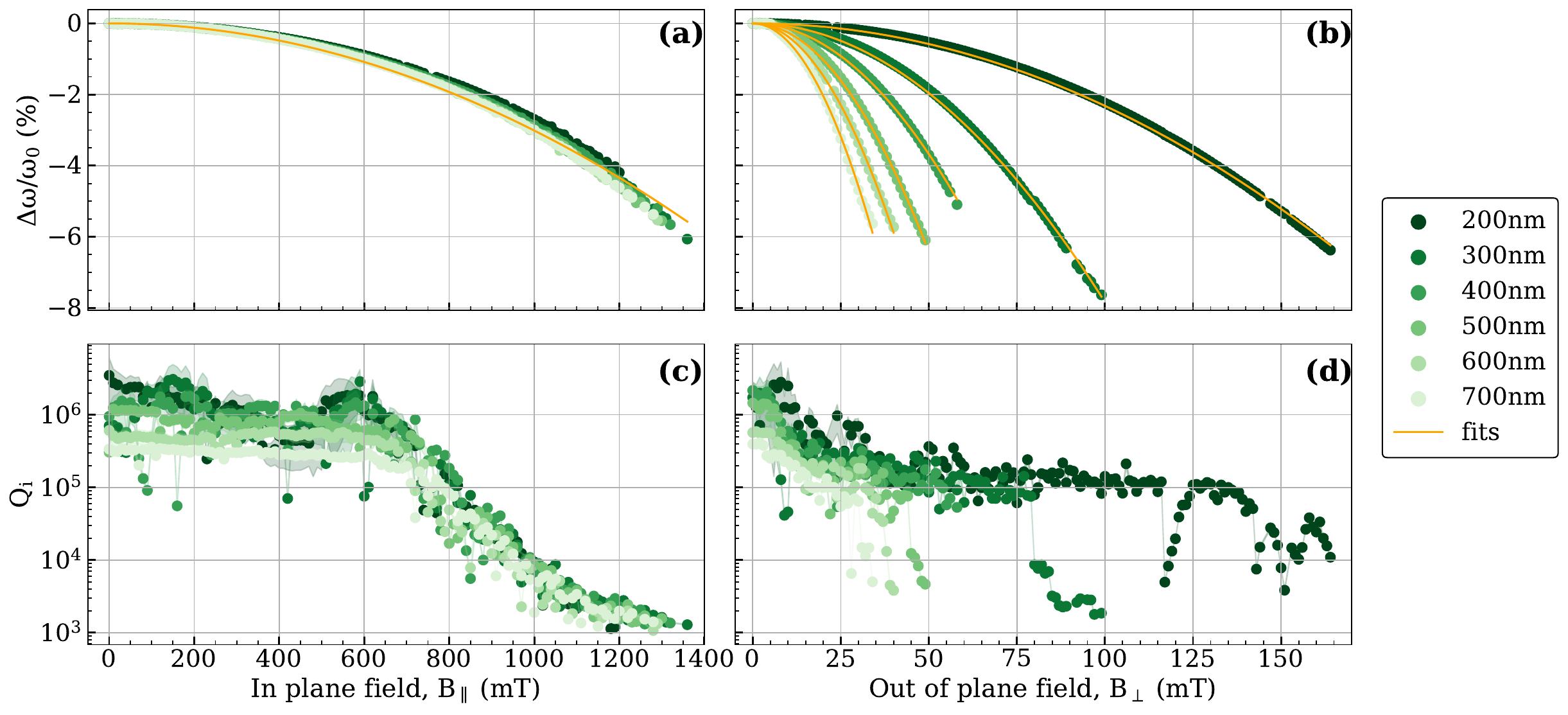}
\caption{\textbf{Magnetic field response of grAl resonators.}  
\textbf{(a)} Normalized resonant frequency shift as a function of in-plane magnetic field. The plot shows the relative frequency shift $\Delta \omega / \omega_0$ for all grAl resonators under an applied in-plane field $B_\parallel$. The fitted curves (orange lines), based on Eq.~\eqref{eq:BC}, allow extraction of the resonators' critical in-plane magnetic fields $B_{C,\parallel}$. The field sweep extends up to 3~T, but the resonant frequencies become indistinguishable before reaching this limit.  
\textbf{(b)} Normalized resonant frequency shift as a function of out-of-plane magnetic field. The extracted resonator critical out-of-plane magnetic fields $B_{C,\perp}$ are determined from the fits (orange lines)  based on Eq.~\eqref{eq:BC}.  
\textbf{(c)} Evolution of the internal quality factor $Q_i$ with increasing in-plane magnetic field $B_\parallel$ for all six grAl resonators.  
\textbf{(d)} Evolution of the internal quality factor $Q_i$ with increasing out-of-plane magnetic field $B_\perp$ for all six grAl resonators (see also Fig.~\ref{fig:gral_bfield_offset}).}
\label{fig:gral_bfield}
\end{figure*}

\subsection{\label{sec:nbn_design}Device design and characterization at optimal conditions}

When aluminum (Al) thin films are deposited in a controlled atmosphere containing a small amount of oxygen, the resulting material consists of highly pure Al grains embedded within an amorphous aluminum oxide matrix~\cite{Deutscher_1973}. This structure contrasts with NbN films, which form a disordered polycrystalline alloy~\cite{dane_bias_2017}. In grAl films, Cooper pairs within the Al grains tunnel through the surrounding oxide barriers, generating a supercurrent that behaves analogously to a network of randomly distributed Josephson junction (JJ) arrays~\cite{maleeva_2018, Cohen_1968}. The effective inductance of the material depends on the degree of oxidation and the grain size~\cite{Levy-Bertrand_2019}. For this study, Al was sputtered in an argon-oxygen (Ar/O$_2$) atmosphere (details in Appendix~\ref{app:materials}). A 50~nm-thick grAl film with $L_k = 149$ pH/sq was selected for this experiment as it provided the closest available match to the NbN film in terms of $L_k$.

A representative resonance spectrum for the 200~nm-wide grAl resonator, along with its corresponding fit based on Eq.~\eqref{eq:S21}, is shown in Fig.~\ref{fig:gral_characteristics}\textbf{(a)}. To gain a more comprehensive understanding of the device characteristics, power scans were conducted for all resonators [see Fig.~\ref{fig:gral_characteristics}\textbf{(b)}]. Due to the highly overcoupled nature of the resonators ($Q_i / Q_c \approx 60$), the extracted values of $Q_i$ in the high-power regime are subject to considerable fitting errors, as small deviations in the fitting curve lead to significant shifts in the extracted parameters. To validate these findings, we fabricated additional resonators from the same grAl film with near-critical coupling. Measurements on these devices [see Fig.~\ref{fig:gral_crit}] show  significantly lower fitting errors and closely matched the trends seen in the overcoupled resonators (for further details, see Appendix~\ref{app:films}).
In the low-photon regime ($\langle n_{ph} \rangle < 1$), the internal quality factor of the grAl resonators ranges from $4 \times 10^5$ to $1.2 \times 10^6$, increasing up to approximately $3 \times 10^6$ at higher intracavity photon numbers ($\langle n_{ph} \rangle > 1$). Notably, a strong dependence of $Q_i$ on resonator width $w$ is observed. In the low-photon regime, the 200~nm-wide resonator exhibits the lowest internal quality factor, $Q_i \sim 2 \times 10^4$; however, at higher photon numbers, it achieves the highest $Q_i \sim 3 \times 10^6$. This behavior stands in contrast to that observed for NbN resonators.

A study of the self-Kerr $K$ of the grAl resonators is performed similarly to the NbN case. The nonlinear response of the resonators is fitted using the same analytic routine (reported in Appendix~\ref{app:selfK-est}), and the extracted values are presented in Fig.~\ref{fig:gral_characteristics}\textbf{(c-d)}. The expected Kerr values can be estimated using Eq.~\eqref{eq:Kerr_BCS} obtained from the BCS theory in the dirty superconductor limit, where the critical currents $I_*$ (ranging from $I_* = 22.3$~$\mu$A for the 200~nm wide resonator to $I_* = 78.1$~$\mu$A for the 700~nm wide resonator) are obtained from a device fabricated on the same grAl film and extracted following the method described in Ref.~\cite{frasca_2024}. For grAl resonators, an alternative approach to estimating the $K$ nonlinearity has recently been proposed~\cite{maleeva_2018}, leveraging their structural similarity to JJ array resonators. Specifically, the granular nature of the material—comprising pure Al grains separated by an amorphous oxide matrix—can be effectively modeled as a series of JJs. This analogy provides a useful framework for understanding the nonlinear properties of grAl resonators and complements the predictions from BCS theory. This leads to the following expression of self-Kerr~\cite{maleeva_2018}:
\begin{equation}
    K = -\frac{3}{16} \pi e a \frac{\omega_r^2}{j_{\text{sw}} V_g},
    \label{eq:Ker_JJ}
\end{equation}
where $a$ is the average grain size, $j_{\text{sw}} = I_{\text{sw}}/wt$ is the switching current density (with $t$ the film thickness), and $V_g = lwt$ is the resonator volume. The switching currents $I_{\text{sw}}$ (ranging from $I_{\text{sw}} = 6.8$~$\mu$A for for the 200~nm wide resonator to $I_{\text{sw}} = 23.8$~$\mu$A for for the 700~nm wide resonator) are extracted from a reference device, following the same procedure as for Ref.~\cite{frasca_2019}.

Estimates of the self-Kerr coefficient $K$, obtained using both the BCS theory and the JJ array model, show similar trends, albeit with a slight offset between them, as illustrated in Fig.~\ref{fig:gral_characteristics}\textbf{(d)}. Determining which model better describes the experimental data is challenging due to possible uncertainties in input-power calibration across different frequencies. Nevertheless, both approaches yield values of $K$ consistent within the same order of magnitude, with maximum deviations from the experimental data of approximately 40\% (8~Hz/photon) for the BCS model and 30\% (13~Hz/photon) for the JJ model.
Interestingly, the JJ array model better captures the fitted $K$ values at lower nonlinearities, whereas the BCS model shows improved agreement at higher $K$ values. This trend, also observed in Ref.~\cite{gupta2024lowlosslumpedelementinductors}, may indicate intrinsic limitations of the JJ array model in fully capturing the nonlinear behavior of grAl resonators.
Overall, the $K$ coefficients for the studied grAl film are an order of magnitude higher than those measured for the NbN film. Despite the similar resonator designs, differences in critical current $I_*$ and $L_k$ ($\sim \times 2$) lead to significant variations in nonlinear properties between the two materials.

\subsection{\label{sec:gral_bfield_chara}Characterization in high B field}

\begin{figure}
\centering
\includegraphics[width=\linewidth]{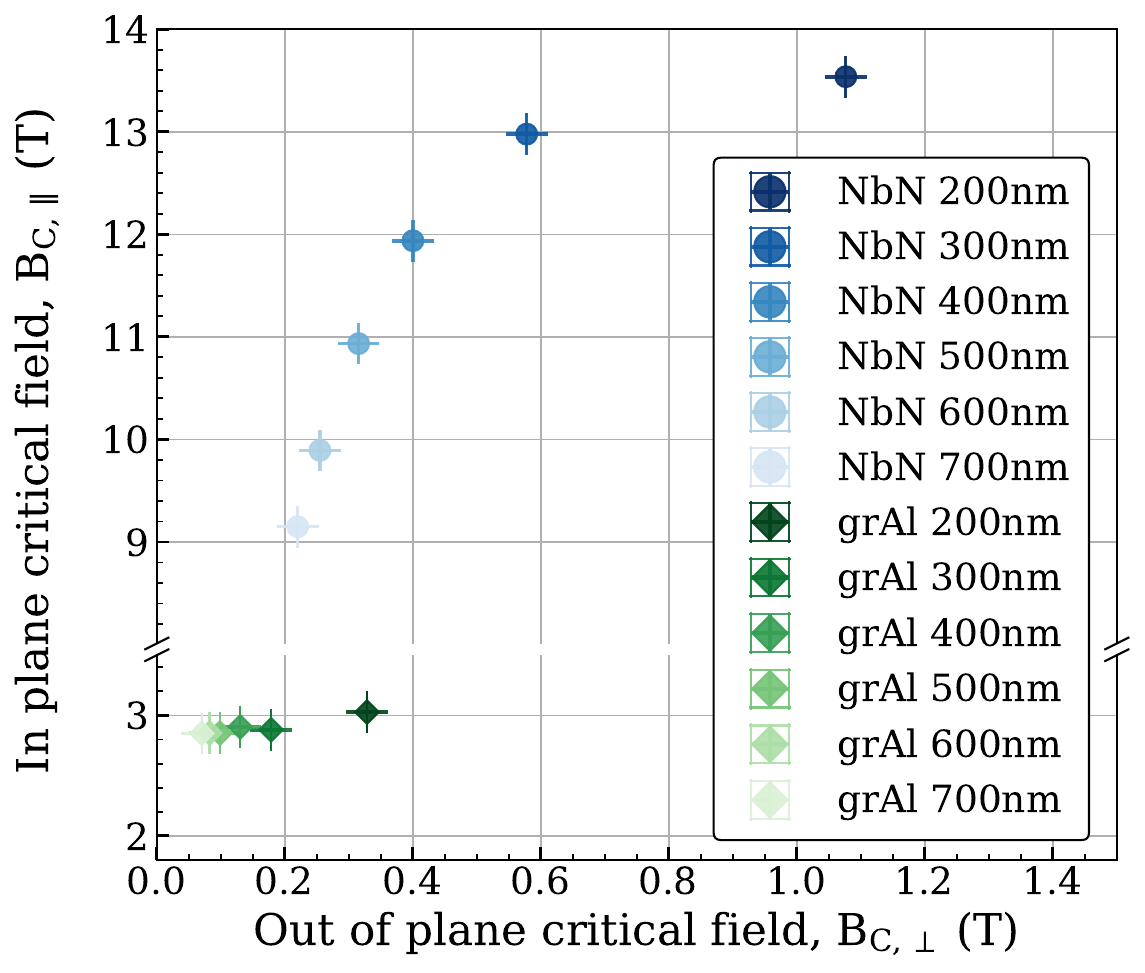}
\caption{\textbf{$B_C$ comparison between NbN and grAl resonators}. Extracted in-plane ($B_{C,\parallel}$) and out-of-plane ($B_{C,\perp}$) critical magnetic fields for NbN (blue) and grAl (green) resonators across all widths according to Eq.~\eqref{eq:BC}.}
\label{fig:bc_comp}
\end{figure}

Following the methodology applied to NbN resonators, the magnetic field characterization of the grAl resonators is conducted for both in-plane and out-of-plane field configurations. The normalized resonator resonant frequency shift $\Delta \omega / \omega_0$ is fitted using Eq.~\eqref{eq:BC} to extract the resonator critical field $B_C$, as shown in Fig.~\ref{fig:gral_bfield}\textbf{(a-b)}. The magnet misalignment is assumed to be the same as in the NbN case (more details in Appendix~\ref{app:bfield_misalignment}). 
For in-plane fields $B_\parallel$, the average critical field is approximately $B_{C,\parallel} \approx 3$~T. Note that, at the applied parallel field of approximately 1.4~T, there is no appreciable difference in $\Delta \omega / \omega_0$ among the six resonators, as shown in Fig.~\ref{fig:gral_bfield}\textbf{(a)}.
For out-of-plane magnetic field $B_\perp$, a strong dependence of both $\Delta \omega / \omega_0$ and $B_{C,\perp}$ on the resonator width $w$ is observed. The maximum out-of-plane critical field reaches $B_{C,\perp} = 330$~mT for the 200~nm wide resonator and $B_{C,\perp} = 70$~mT for the 700~nm wide resonator.

Although the critical fields of grAl resonators are lower than those of NbN resonators—primarily due to the smaller superconducting gap of grAl—the evolution of $Q_i$ under increasing magnetic field exhibits distinct differences [see Fig.~\ref{fig:gral_bfield}\textbf{(c)}]. For in-plane fields, grAl resonators maintain a nearly constant $Q_i$ up to approximately 600~mT (the "cutoff" field for the given film thickness), beyond which a sharp drop occurs. Unlike in NbN resonators, no clear evidence of ESR coupling with paramagnetic spin impurities is observed, a behavior consistent with previous observations in Ref.~\cite{Janik_2025}.

Out-of-plane field measurements exhibit a similar trend: after an initial slight decrease, $Q_i$ remains relatively stable up to a characteristic "cutoff" field, which depends on the resonator width $w$ [see Fig.~\ref{fig:gral_bfield}\textbf{(d)} and Fig.~\ref{fig:gral_bfield_offset} in Appendix~\ref{app:bfield_perp_offset}]. A maximum cutoff field of 140~mT is observed for the 200~nm-wide resonator. Small fluctuations in $Q_i$ throughout the magnetic field sweep are attributed to vortex-induced dissipation, which becomes more pronounced as the applied field increases.

\section{Discussion and Conclusion}

This study presents a detailed comparison of high-kinetic inductance superconducting resonators fabricated from NbN and grAl thin films (of thickness 13~nm and 50~nm, respectively), focusing on their distinct strengths and limitations for quantum technologies and hybrid device applications.

At zero magnetic field, all grAl resonators exhibit higher internal quality factors ($Q_i$) than their NbN counterparts. While lower-$L_k$ NbN films have been shown to achieve higher $Q_i$ values~\cite{frasca_2023}, reducing the kinetic inductance in our study would have compromised the direct comparison of the two materials under similar $L_k$ conditions. In contrast, the high $Q_i$ of grAl resonators suggests the potential to further increase $L_k$ without sacrificing quality. It is worth noting that the chosen sheet kinetic inductance $L_k \approx 100$~pH/sq lies at the lower end of values typically reported for grAl films, yet approaches the upper limit of what can be achieved with NbTiN and NbN films~\cite{Glezer_Moshe_2020_nbn_vs_gral}.
Increasing $L_k$ further generally comes at the cost of a lower $T_C$ and reduced film homogeneity across the wafer~\cite{Levy-Bertrand_2019}.

In terms of magnetic field resilience, NbN outperforms grAl, with measured in-plane critical fields ($B_{C,\parallel}$) reaching up to 13.5~T and out-of-plane fields ($B_{C,\perp}$) up to 1.07~T for the 200~nm wide resonator. The evolution of the NbN resonator losses under applied in-plane fields exhibits three distinct regions: an initial electron spin resonance dip around 300~mT, followed by a "sweet spot" near 1~T, where $Q_i$ improves by an order of magnitude relative to the zero-field case. This enhancement is attributed to an increased frequency detuning of magnetic-field-sensitive TLS away from the resonator's resonant frequency~\cite{Bahr_2024}, although further investigation is needed to fully elucidate the underlying mechanism.
From a technological standpoint, this phenomenon could be explored to enhance device performance without the need to saturate TLSs through high-photon-number operation or introducing additional proximity pumping tones near the resonator’s resonant frequency~\cite{Andersson_2021}.
Beyond 1~T, $Q_i$ decreases exponentially with increasing field strength. Nevertheless, even at 6~T in-plane magnetic field, the 200~nm-wide NbN resonator retains a $Q_i$ above $5 \times 10^4$, which remains sufficiently high for several applications in the hybrid circuit QED context.

GrAl, while less resilient to magnetic fields—achieving, in our case, $B_{C,\parallel} \approx 3$~T and $B_{C,\perp} \approx 0.33$~T for the 200~nm wide resonator—demonstrates robustness in $Q_i$ up to a well-defined "cutoff" field. This cutoff field depends on the resonator cross-section perpendicular to the applied magnetic field, reaching approximately 600~mT for in-plane fields and 150~mT for out-of-plane fields in our devices~\cite{Janik_2025}. Unlike the NbN case, grAl resonators exhibit no evidence of an ESR dip, making them promising candidates for field-sensitive applications for which suppression of spurious spin signals is necessary~\cite{Bahr_2024}. A comparison of the extracted $B_C$ values for both materials is shown in Fig.~\ref{fig:bc_comp}. While the out-of-plane resilience of grAl resonators is lower than that of NbN, the overall $B_C$ values remain sufficiently high to meet the requirements of certain hybrid quantum technology applications~\cite{Jirovec_2021}.

The observed nonlinearity in grAl resonators exceeds that in NbN by over an order of magnitude for comparable resonator geometries. Self-Kerr $K$ interactions in the MHz range have been already demonstrated in high-aspect-ratio weak links fabricated from high-$L_k$ grAl thin films~\cite{winkel_2020,Rieger_2022}. Both materials exhibit nonlinear behavior consistent with theoretical expectations, reinforcing their reliability for device engineering.

Ultimately, the choice between NbN and grAl depends on the specific application and required performance metrics. For high-field resilience above 1~T, where moderate nonlinearity suffices, NbN is an excellent candidate, particularly for magnetic field and temperature resilient resonators and quantum limited parametric amplifiers~\cite{Xu_2023,frasca_2024}. In contrast, grAl is more suitable for applications operating below the "cutoff" field (approximately 1~T in our case), where maximizing nonlinearity and impedance is critical, such as in magnetic field-compatible devices designed for strong coupling to spin qubits~\cite{De_Palma_2024}. However, the fabrication of grAl films still presents additional challenges due to their lower reproducibility~\cite{Janik_2025}.

\section*{Contributions}

S.F. and P.S. conceived the experiments. S.F. developed the recipes and optimized the deposition and characterization techniques. C.R and S.F. fabricated the device. C.R. performed measurements. C.R. analyzed the data with inputs from S.F. and P.S. C.R., S.F. and P.S. wrote the manuscript. P.S. supervised the work.

\section*{Acknowledgments}

This research was partly supported by the Swiss National Science Foundation (SNSF) through the grants Ref. No. UeM019-16 – 215928, Ref. No. 200021\_200418 and Ref. No. 206021\_205335. 
We also acknowledge the support from the Swiss State Secretariat for Education, Research and Innovation (SERI) under contract number 01042765 SEFRI MB22.00081 and the support from the NCCR Spin Qubit in Silicon (NCCR-SPIN) Grant No. 51NF40-180604.
S.F. acknowledges the support of SNF Spark project 221051 and of the EPFL Quantum Innogrant. We also acknowledge the Center of Microtechnology (CMi, EPFL) for the access and maintenance of the cleanroom without which fabrication wouldn't be possible.

\appendix

\section{\label{app:materials}Materials}

\begin{figure}
\centering
\includegraphics[width=\linewidth]{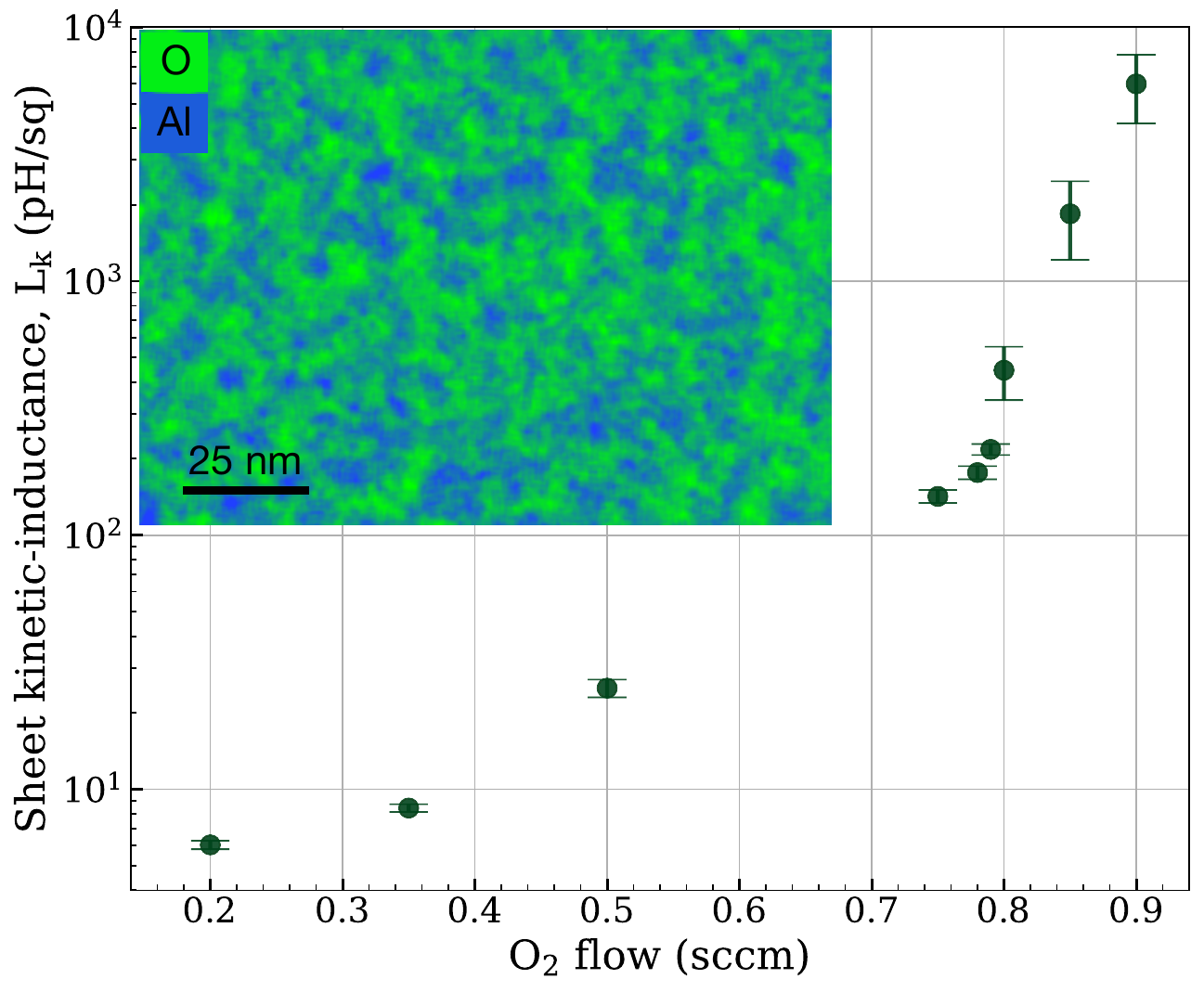}
\caption{\textbf{grAl $L_k$ with O$_2$ flow}. Variation of $L_k$ for the deposited grAl film as a function of increasing O$_2$ flow in the sputtering chamber (keeping a constant deposition time). Sheet resistance measurements, obtained via a 4-probe method, provide the resistivity for each film at room temperature (see Eq.~\eqref{eq:Lk_withR}). The inset shows an energy-dispersive X-ray (EDX) analysis, highlighting the film's structure with aluminum grains ($\approx$ 6~nm in diameter) distributed within an oxide matrix.}
\label{fig:gral_flow}
\end{figure}

\subsection{\label{app:films}Films Deposition}

Both NbN and grAl films were deposited at room temperature using a Kenosistec RF sputtering system. 
The 13~nm-thick NbN film was fabricated via bias sputtering~\cite{dane_bias_2017}, a technique that promotes the formation of polycrystalline films, improving device yield. The film’s kinetic inductance ($L_k$) was estimated by measuring its room-temperature sheet resistance using a four-probe system. For Nb sputtered at a chamber pressure of 5~$\mu$bar with 8\% nitrogen in an argon atmosphere over 8 minutes and 30 seconds, the resulting sheet resistance was 220~$\Omega$/sq, corresponding to $L_k \approx 90$~pH/sq (further details in Appendix~\ref{app:Lk_est}). To achieve a desired $L_k$, the film thickness was adjusted by varying the sputtering duration, while maintaining fixed sputtering parameters.

The 50~nm-thick grAl film was fabricated by sputtering aluminum in an atmosphere containing a controlled partial pressure of O$_2$ mixed with Ar. 
The sputtering parameters—150~W power, 5~$\mu$bar pressure, and a sputtering duration of 5 minutes—were kept constant, while the O$_2$ flow rate was varied to tune the film’s $L_k$, as shown in Fig.~\ref{fig:gral_flow}. The data reveals an exponential dependence of $L_k$ on oxygen flow. However, particularly near the inflection point, film reproducibility remains limited, as slight parameter fluctuations strongly affect the film’s properties. 
Fig.~\ref{fig:gral_flow} also includes a TEM-EDX measurement of the grAl film, confirming its granular structure, with Al grains of approximately 6~nm in diameter embedded within a surrounding oxide matrix.

\subsection{\label{app:fab}Device fabrication}

\begin{figure}
\centering
\includegraphics[width=\linewidth]{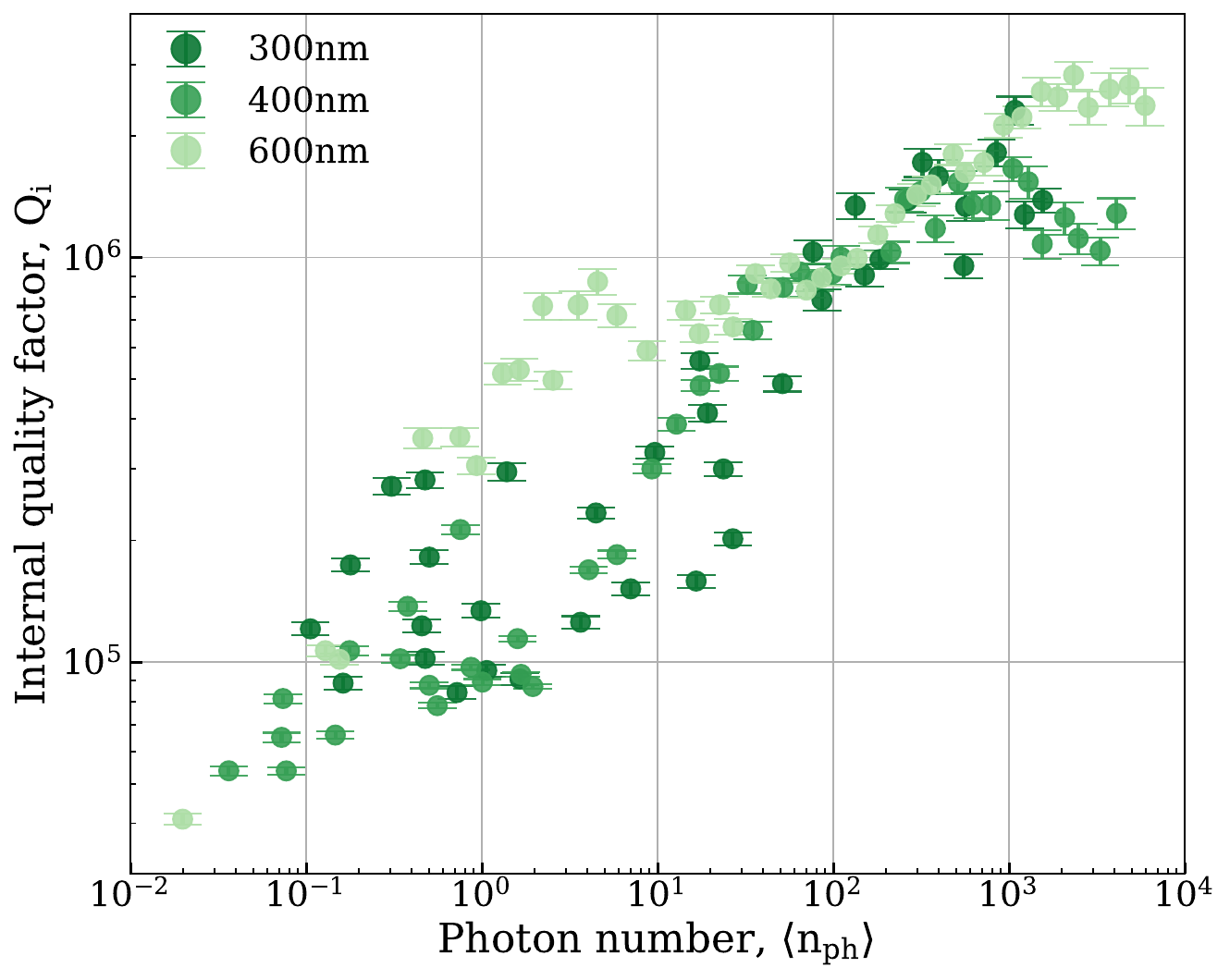}
\caption{\textbf{grAl resonator close-to-critical coupling}. Internal quality factor $Q_i$ evolution on average intracavity photon number ($\langle n_{ph} \rangle$) for three grAl resonators designed to be close to the critically coupled regime ($Q_i \sim Q_c$). These devices were fabricated from the same film as the overcoupled resonators reported in Fig.~\ref{fig:gral_characteristics}, maintaining identical geometries but with increased distance from the feedline.}
\label{fig:gral_crit}
\end{figure}

The fabrication process for NbN devices begins with a surface cleaning procedure to remove native oxides and contaminants. A high-resistivity ($\rho \geq 10~k\Omega \cdot$cm) intrinsic 4-inch silicon wafer with $\braket{100}$ crystalline orientation is immersed in a piranha bath (heated to 100$^\circ$C) for 10 minutes, followed by a 2-minute dip in a 40\% HF solution. 
Following surface preparation, a 13~nm-thick NbN film is deposited via bias sputtering~\cite{dane_bias_2017} at room temperature using a Kenosistec RF sputtering system. The wafer is then diced into 11 × 16~mm chips. Each chip is spin-coated with a 100~nm-thick layer of ZEP-502A 50\% positive e-beam resist at 4500~rpm, followed by a baking step at 150$^\circ$C for 5 minutes. Device patterns are defined via electron beam lithography (Raith EBPG5000+ at 100~keV) and developed in n-amyl acetate for 1 minute, followed by a rinse in a 9:1 mixture of methyl isobutyl ketone (MiBK) and isopropanol (IPA) for 1 minute. The NbN film is then patterned via reactive ion etching (RIE) using a CF$_4$/Ar gas mixture at 15~W power for 5 minutes. Residual resist is removed by immersing the sample in heated Microposit Remover 1165 (70$^\circ$C), followed by two cleaning steps with sonication in acetone and IPA. After processing, the wafer is coated with a 4~$\mu$m-thick layer of AZ P4K-AP protective resist for device protection. Finally, the wafer is diced into 4 × 7~mm chips, which are ready for packaging and subsequent measurement.

The fabrication of grAl devices follows a similar procedure, with two notable differences. First, a 400~nm-thick layer of ZEP-502A 100\% positive resist is spin-coated at 3000~rpm for e-beam lithography. Second, the etching step employs a Cl$_2$/BCl$_3$-based chemistry in an inductively coupled plasma (ICP) etcher (800~W coil power, 50~W RF power, chamber pressure of 3~mTorr) for 1 minute, resulting in approximately 10~nm of overetching into the Si substrate.

\subsection{\label{app:Lk_est}$L_k$ estimation}

\begin{figure}
\centering
\includegraphics[width=\linewidth]{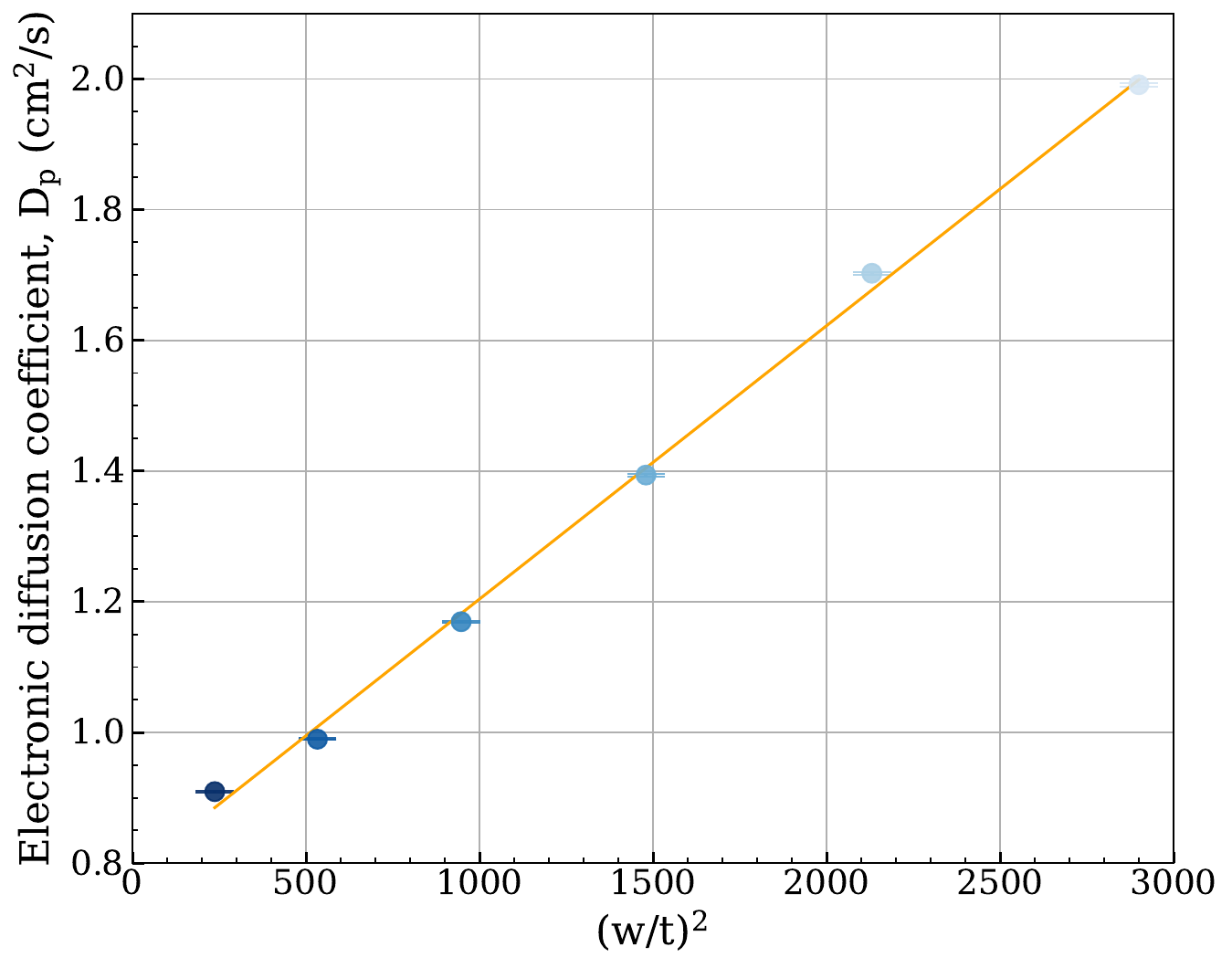}
\caption{\textbf{Magnet misalignment fitting}. Generalized electron diffusion coefficient $D_p$ plotted against the square of width-to-thickness ratio ($w^2/t^2$) for each NbN resonator. Using Eq.~\eqref{eq:freq_shift_bfield_inplane}, the generalized coefficient is expressed as $D_p = D(1+\theta_B^2 w^2/t^2)$, allowing $D_p$ to be fitted for each resonator. From this fit, the magnet misalignment angle $\theta_B$ is estimated to be approximately 1 degree.}
\label{fig:nbn_misalignement}
\end{figure}

\begin{figure}[h!]
\centering
\includegraphics[width=\linewidth]{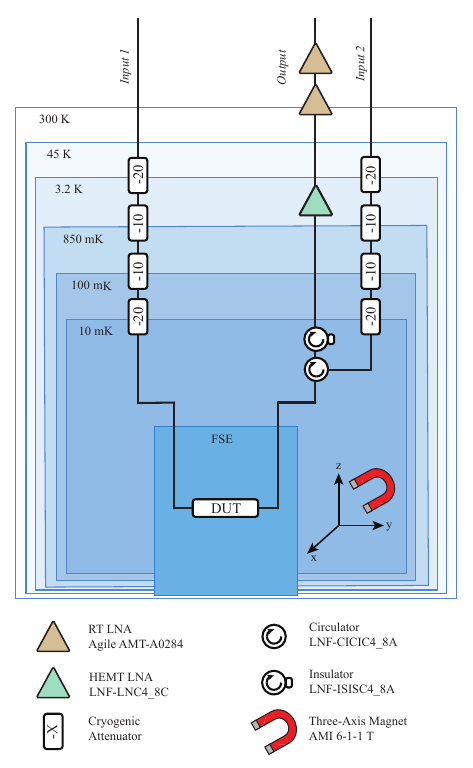}
\caption{\textbf{Cryogenic Setup}. Experimental setup for measurements conducted in an LD Blufors cryogenic system with bottom-loading capabilities. The transmission input line (Input 1) includes 60 dB of cold attenuation, with additional room-temperature attenuation applied as needed for specific measurements. The total attenuation of the input lines are characterized at room temperature to extract the input power for self-Kerr estimation. A 6-1-1~T vector magnet, is mounted at the 4 K stage. A secondary input line (Input 2) is available for reflection measurements. After interaction with the device under test (DUT), the signal passes through a circulator and an isolator before being amplified by a HEMT amplifier at the output line, followed by one or two extra amplification stages at room temperature.}
\label{fig:exp_setup}
\end{figure}

Following sputtering, the sheet resistance $R_{sq}$ of the films is measured using a four-probe system (AIT CMT-SR2000N Resistivity Measurement System). The kinetic inductance ($L_k$) can then be estimated using the relation~\cite{tinkham2004introduction}:

\begin{equation}
    L_k = \frac{R_{sq} \hbar}{\pi \Delta} \frac{1}{\tanh[\Delta/(2k_B T)]},
    \label{eq:Lk_withR}
\end{equation}
where $\Delta \approx 1.764k_B T_C$ is the superconducting gap and $T_C$ is the critical temperature of the film. 

The exact $T_C$ of a newly sputtered film is typically unknown, making it challenging to directly determine $L_k$ from the room-temperature resistance measurement alone. The most reliable method for determining $L_k$ involves fabricating test resonators from the same film. These resonators are initially designed to operate at specific frequencies based on an estimated $L_k$.

\begin{figure*}
\centering
\includegraphics[width=.96\linewidth]{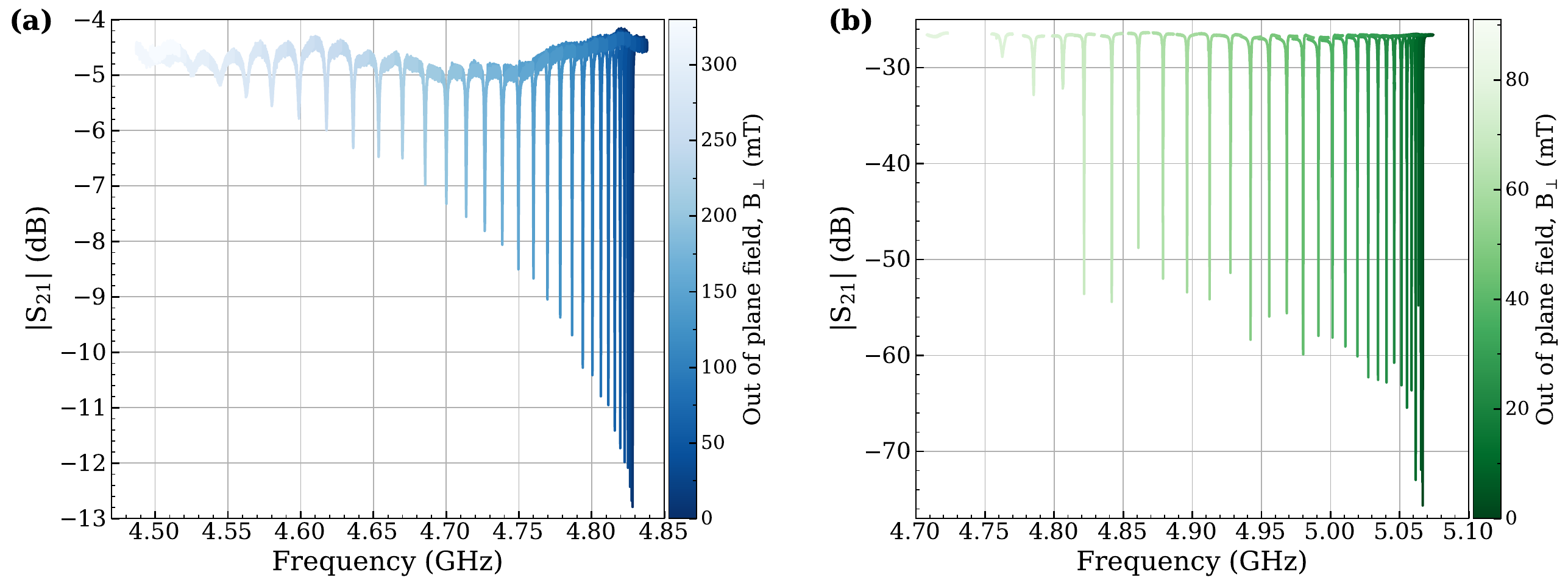}
\caption{\textbf{Spectroscopy evolution in out-of-plane magnetic field}. \textbf{(a)} Amplitude ($|S_{21}|$) for the NbN 300~nm wide wire resonator recorded at different values of out-of-plane magnetic field ($B_\perp$) and displayed on the same plot. The color bar on the right represents the $B_\perp$ intensity. A clear resonator resonant frequency shift is observed with by a continuous exponential reduction in resonance depth, indicating a reduction in $Q_i$ as $B_\perp$ is increased. \textbf{(b)} Amplitude ($|S_{21}|$) for the grAl 300~nm wide wire resonator recorded at different values of out-of-plane magnetic field ($B_\perp$) and displayed on the same plot. The color bar on the right represents the $B_\perp$ intensity. Unlike the NbN device, the grAl resonator exhibits less pronounced continuous reductions in resonance depth, showing instead abrupt changes approaching the "cutoff" field (around 80 mT in this case).}
\label{fig:bfield_spectros}
\end{figure*}

To predict the resonator frequencies, simulations are performed using Sonnet software, following the procedure in~\cite{frasca_2023}. After fabrication and measurement, the experimentally observed resonator frequencies are compared to the simulated values. In Sonnet, a sweep of $L_k$ values is conducted, providing a range of potential resonance frequencies for the resonator under study. 
These resonant frequencies are then fitted using the relation for $\lambda/4$ resonators: 
\begin{equation} f = \frac{1}{4 l \sqrt{\tilde{L} \tilde{C}}}, \label{eq:freq_master} 
\end{equation} where $l$ is the resonator length, and $\tilde{L}$ and $\tilde{C}$ are the inductance and capacitance per unit length, respectively. Here, $\tilde{C}$ is treated as a fitting parameter.
By comparing the fitted frequencies from Sonnet simulations with the measured resonator frequencies, the $L_k$ of the film can be accurately estimated. Once $L_k$ is known, Eq.~\eqref{eq:Lk_withR} can be used to determine the critical temperature $T_C$ of the film, since both the resistance and $L_k$ are now available. This method provides a reliable reference for predicting $L_k$ in films without requiring additional test resonators.
Applying this approach to the grAl film, a critical temperature of $T_C = 2.1$~K is obtained, while for the NbN film, $T_C \approx 4$~K, consistent with previous findings~\cite{frasca_2023}. Additionally, this method proves useful for estimating other key parameters, such as the characteristic impedance of the resonators.

\section{\label{app:methods}Methods}

\subsection{\label{app:setup}Experimental Setup}

The measurements were performed using a LD250 BlueFors dilution cryostat equipped with a Fast Sample Exchange (FSE) system. The device was positioned below the mixing chamber (MXC) plate at a base temperature of 7~mK. The cryostat includes a superconducting magnet (American Magnetics 6-1-1~T Vector Magnet) located at the 4~K stage. The samples were mounted in the FSE using a copper plate and brass screws, ensuring positioning at the region of maximum magnetic field intensity. 
Signal input and output were handled by a two-port Vector Network Analyzer (VNA, Rohde \& Schwarz ZNB series) to acquire the resonator scattering parameters. The input signal was attenuated by a total of 80–100~dB (depending on the measurement conditions), including 20–40~dB attenuator at room temperature (BlueFors cryo attenuator), 40~dB attenuation along the cryostat stages down to the MXC, and an additional 20~dB attenuator at MXC. 
After interacting with the device, which is designed in a hanger configuration, the output signal passed through a circulator and an isolator before being amplified. The first stage of amplification was performed by a high-electron-mobility transistor (HEMT, LNF-LNC4\_8C) at the 4~K stage, followed by one or two room-temperature amplifiers (Agile AMT-A0284). A schematic of the experimental setup is shown in Fig.~\ref{fig:exp_setup}.

For reference measurements of the critically coupled grAl resonators [see Fig.~\ref{fig:gral_crit}], the same room-temperature setup was used. However, the dilution cryostat was an LD250 BlueFors system without a bottom loader, and the samples were mounted in a homemade cold finger. Additional infrared-absorbing eccosorb filters (Quantum Microwave CRYOIRF-003MF-S) were installed on both the input and output lines at the MXC stage. A 4–8 GHz band-pass filter (MiniCircuits ZBSS-6G-S+) was added in the output at MXC level to reduce the effect of potential noise coming from the HEMT.

Regarding chip packaging, the chips were glued to a copper mount using PMMA, which was heated to 90$^\circ$C for 15 minutes. A gold-plated PCB was then screwed onto the copper mount, and electrical connections between the PCB and chip were realized via 25~$\mu$m-wide aluminum wire bonding.

\subsection{\label{app:spectroscopy_fit}Spectroscopy fitting}

\begin{figure}
\centering
\includegraphics[width=\linewidth]{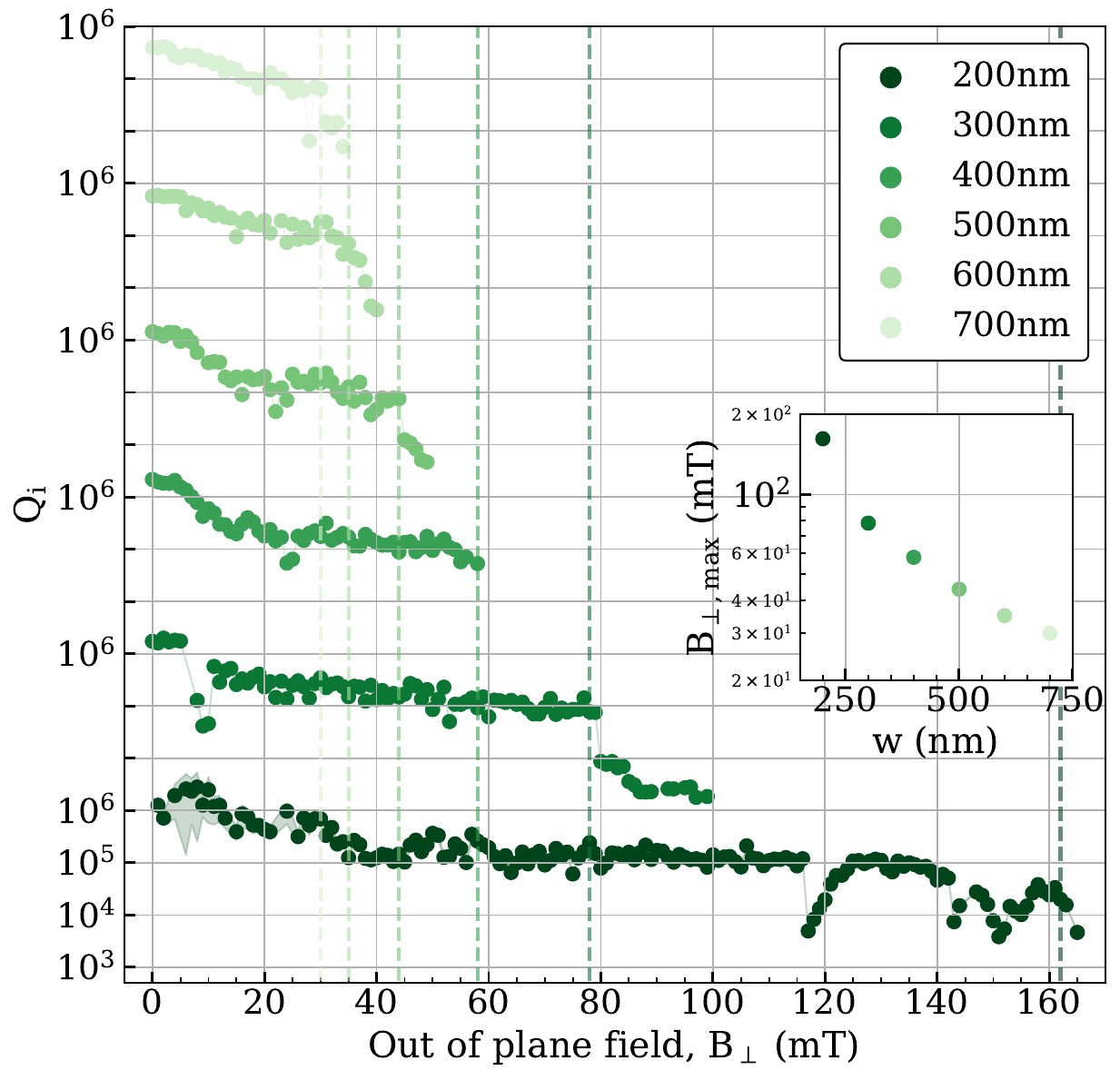}
\caption{\textbf{Evolution of grAl resonator $Q_i$ with $B_{\perp}$}. Evolution of the internal quality factor $Q_i$ with respect to out-of-plane magnetic field $B_{\perp}$ with an artificial $10^3$ offset step introduced between each dataset. The vertical dashed lines are indicates the $B_{\perp,\text{max}}$ (previously referred to as the "cutoff" field in the main text). These values are plotted in inset with respect to the resonator width $w$.}
\label{fig:gral_bfield_offset}
\end{figure}

Experimental data acquired with the VNA were fitted using a master equation model for the S-parameters [Eq.~\eqref{eq:S21}]. The feedline transmission $S_{21}$ around each of the resonator resonant frequency was first fitted using the Probst library \texttt{resonator tools}~\cite{Probst_2015}, which provided accurate initial guesses for the fitting function~\cite{chen_2022}:
\begin{equation}
    S_{21} = ae^{i\alpha}e^{-2\pi i f_d \tau} \bigg(1- \frac{\kappa}{\kappa + \gamma} \frac{e^{i \phi}}{\cos(\phi)} \frac{1}{1+2i \frac{\Delta_r}{\kappa + \gamma} } \bigg),
    \label{eq:S21}
\end{equation}
where $\Delta_r = \omega_d - \omega_0$ is the detuning between the drive frequency $\omega_d$ and the resonator resonant frequency $\omega_0$, while $\kappa$ and $\gamma$ represent the resonator external and internal loss rates, respectively. The parameters $a$, $\alpha$, $\tau$, and $\phi$ account for environmental corrections, such as impedance mismatches.
It is important to note that this fitting approach does not capture nonlinear effects of the resonators, as discussed in Appendix~\ref{app:selfK-est}. 

A second-stage fitting was performed using the Python library \texttt{Non-Linear Least-Squares Minimization and Curve-Fitting} (LmFit), with initial values provided by the \texttt{resonator tools} method. In some cases, fitting failures occurred due to factors such as overcoupling, standing waves, or measurement noise. To ensure clarity and accuracy in the presented results, datasets with fitting errors exceeding $10^3$ were excluded from the analysis.

For power scans, the average intracavity photon number ($\langle n_{ph} \rangle$) was estimated following the relation~\cite{Probst_2015}:

\begin{equation}
    \langle n_{ph} \rangle = C \frac{\kappa}{\hbar \omega_0 (\kappa + \gamma)^2} 10^{\frac{P-A}{10}}/1000,
    \label{eq:nph_eq}
\end{equation}
where $C$ is a configuration-dependent factor ($C = 4$ for hanger configuration), $P$ is the VNA output power in dB, and $A$ is the total attenuation from the VNA to the device under test. The total attenuation $A$ was estimated by measuring the input lines at room temperature.

\subsection{\label{app:selfK-est}Self Kerr estimation}

\begin{figure*}
\centering
\includegraphics[width=.96\linewidth]{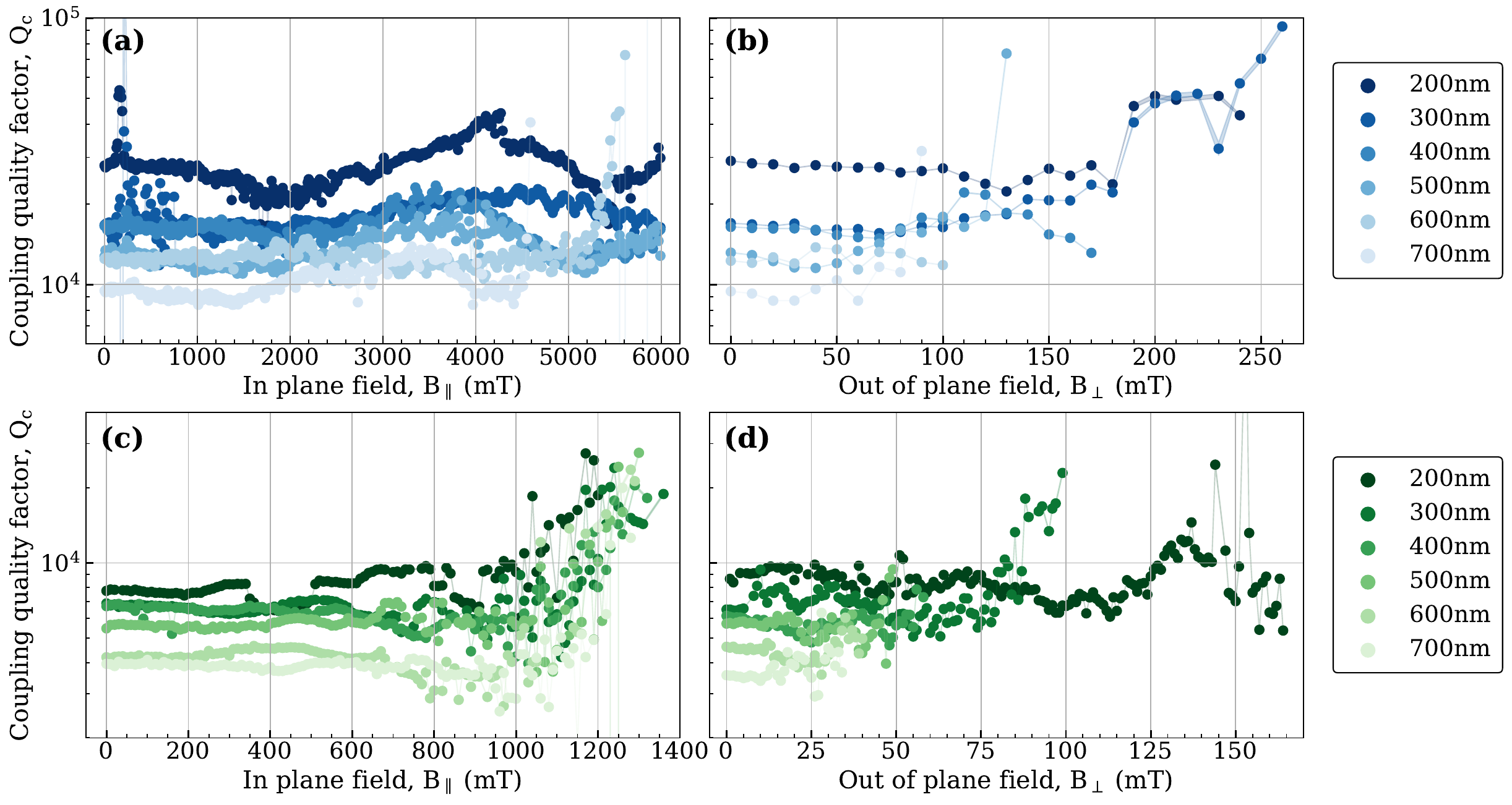}
\caption{\textbf{Evolution of resonator $Q_c$ with magnetic field}. \textbf{(a)} Evolution of the coupling quality factor ($Q_c$) for the NbN resonators during the in-plane magnetic field ($B_\parallel$) sweep. \textbf{(b)} Evolution of $Q_c$ for the NbN resonators during the out-of-plane magnetic field sweep ($B_\perp$). \textbf{(c)} Evolution of $Q_c$ for the grAl resonators during the $B_\parallel$ sweep. \textbf{(d)} Evolution of $Q_c$ for the grAl resonators during the $B_\perp$ sweep. Across all cases, $Q_c$ remains largely stable with increasing magnetic field.}
\label{fig:bfield_qc}
\end{figure*}

The nonlinear dependence of the kinetic inductance $L_k$ on the current $I$ though the resonator (see Eq.~\eqref{eq:Lk_withI}) leads to an effective self-Kerr term, $K$, which can be expressed in the Hamiltonian formalism as:

\begin{equation}
    \hat{H} = \hbar \omega_r \hat{a}^{\dag}\hat{a} + \hbar \frac{K}{2} (\hat{a}^{\dag} \hat{a})^2,
    \label{eq:Hamiltonian_nonlin}
\end{equation}
where $\hat{a}$ is the conventional ladder operator.
Following input-output theory~\cite{eichler_2014, anferov_2020}, the S-parameter model for hanger-type resonators, taking into account nonlinearity, is given by:

\begin{equation}
    S_{21} = 1- \frac{\kappa}{\kappa + \gamma} \frac{e^{i \phi}}{\cos(\phi)} \frac{1}{1+2i(\delta - \xi n)},
    \label{eq:S21_nonlin}
\end{equation}
where 
\begin{equation}
    \delta \equiv \frac{\omega_\text{d}-\omega_0}{\kappa+\gamma}, \quad \xi \equiv \frac{|\tilde{\alpha}_{in}|^2 K}{\kappa+\gamma}, \quad n \equiv \frac{|\alpha|^2}{|\tilde{\alpha}_{in}|^2}.
\end{equation}
$\phi$ is the parameter taking into account impedance mismatches, $\tilde{\alpha}_{in} \equiv \sqrt{\kappa}\alpha_{in}/(\kappa+\gamma)$, $\kappa \equiv \omega_0 / Q_\text{c}$ is the coupling to the feedline, and $\gamma \equiv \omega_0 / Q_\text{i}$ the internal resonator losses. The photon number $n$, representing the normalized number of photons in the resonator, is computed by solving the equation:
\begin{equation}
    \frac{1}{2} = \bigg(\delta^2 + \frac{1}{4}\bigg)n-2\delta \xi n^2 + \xi^2 n^3.
    \label{eq:nonlin_nph}
\end{equation}

To accurately extract the self-Kerr coefficient $K$, we perform a two-dimensional fit of the spectroscopy data of the complex $S_{21}$ as a function of probe frequency and input power delivered to the device, using Eq.~\eqref{eq:S21_nonlin}. The fitting is restricted to data collected below the bifurcation threshold. During the fitting procedure, the parameters $\phi$ and $K$ are optimized, while the decay rates $\kappa$ and $\gamma$ are fixed to the values obtained from fitting Eq.~\eqref{eq:S21} to data taken immediately before significant resonant frequency shifts occur. At each iteration of the fitting routine, Eq.~\eqref{eq:nonlin_nph} is solved numerically to determine $n$.
Accurate knowledge of the power delivered to the device is crucial for reliable $K$ extraction. In this study, the total attenuation to the device was estimated by measuring the input line at room temperature, assuming minimal discrepancies between room-temperature and cryogenic conditions (see Eq.~\eqref{eq:nph_eq}).

The $K$ values are compared to theoretical estimates obtained from Eq.~\eqref{eq:Kerr_BCS}. The critical current $I_*$ and switching current $I_{\text{sw}}$ are determined from previously characterized devices fabricated on the same film (following the method in~\cite{frasca_2024}) and scaled according to the specific resonator geometries.
By substituting $L_t = L_k (l/w)$ and $I_*^2 = j_C^2 t^2 w^2$ into Eq.~\eqref{eq:Kerr_BCS}, we obtain:
\begin{equation}
    K = -\frac{3}{8} \frac{\hbar \omega_r^2}{L_k j_C^2 t^2 (lw)}
    \label{eq:Lk_with_lw}.
\end{equation}
The expected linear dependence of $K$ on $1/(wl)$ becomes evident, which is used as the x-axis in Fig.~\ref{fig:nbn_characteristics}\textbf{(d)} and Fig.~\ref{fig:gral_characteristics}\textbf{(d)}. A similar trend can be deduced by replacing $V_g = twl$ in Eq.~\eqref{eq:Ker_JJ}.

\subsection{\label{app:bfield_sweep_meas}Magnetic field sweep}

\begin{center}
\begin{table*}
\centering
\begin{tabular}{|c|c|c|c|c|c|c|c|c|}
\hline
  & $w$ (nm) & $f_0$ (GHz) & $Q_i$ at $\langle n_{ph} \rangle \approx 1$ & $Q_c$ & $Z$ (k$\Omega$) & $K/2\pi$ (Hz/photon) & $B_{C,\parallel}$ (T) & $B_{C,\perp}$ (mT) \\ \hline
Res 0 & $200$ & $4.0743$ & $(13805 \pm 286)$ & $(28241 \pm 104)$ & 2.725 & $(-4.506 \pm 0.007)$ & $(13.537 \pm 0.009)$ & $(1076.6 \pm 0.6)$ \\ \hline
Res 1 & $300$ & $4.8282$ & $(20344 \pm 366)$ & $(17156 \pm 54)$ & 2.225 & $(-3.877 \pm 0.007)$ & $(12.977 \pm 0.009)$ & $(577.9 \pm 1.1)$ \\ \hline
Res 2 & $400$ & $5.4080$ & $(17324 \pm 248)$ & $(14008 \pm 46)$ & 1.927 & $(-2.690 \pm 0.007)$ & $(11.938 \pm 0.008)$ & $(400.1 \pm 0.9)$ \\ \hline
Res 3 & $500$ & $5.7831$ & $(28542 \pm 627)$ & $(14264 \pm 51)$ & 1.724 & $(-2.479 \pm 0.006)$ & $(10.934 \pm 0.008)$ & $(315.3 \pm 0.9)$ \\ \hline
Res 4 & $600$ & $6.2537$ & $(25832 \pm 431)$ & $(12617 \pm 26)$ & 1.582 & $(-2.323 \pm 0.006)$ & $(9.893 \pm 0.006)$ & $(254.9 \pm 0.9)$ \\ \hline
Res 5 & $700$ & $6.6245$ & $(30575 \pm 692)$ & $(10316 \pm 27)$ & 1.457 & $(-1.999 \pm 0.005)$ & $(9.148 \pm 0.006)$ & $(220.1 \pm 0.6)$ \\ \hline
\end{tabular}
\caption{Summary table reporting the main parameters extracted for of all the NbN resonators with $w$ the resonator width, $f_0$ the resonant frequency, $Q_i$ the internal quality factor, $\langle n_{ph} \rangle$ the average intracavity photon number, $Q_c$ the coupling quality factor, $Z$ the characteristic impedance, $K$ the self-Kerr, and $B_{C,\parallel}$, $B_{C,\perp}$ the in-plane and out-of-plane critical magnetic field respectively.}
\label{table:NbN_resonators}
\end{table*}
\end{center}

\begin{center}
\begin{table*}
\centering
\begin{tabular}{|c|c|c|c|c|c|c|c|c|}
\hline
  & $w$ (nm) & $f_0$ (GHz) & $Q_i$ at $\langle n_{ph} \rangle \approx 1$ & $Q_c$ & $Z$ (k$\Omega$) & $K/2\pi$ (Hz/photon) & $B_{C,\parallel}$ (T) & $B_{C,\perp}$ (mT) \\ \hline
Res 0 & $200$ & $4.2809$ & $(6.75 \pm 1.16) \times 10^5$ & $(8893 \pm 16)$ & 3.530 & $(-49.999 \pm 0.009)$ & $(3.028 \pm 0.007)$ & $(328.4 \pm 0.3)$ \\ \hline
Res 1 & $300$ & $5.0666$ & $(12.65 \pm 3.32) \times 10^5$ & $(6545 \pm 7)$ & 2.882 & $(-44.198 \pm 0.005)$ & $(2.881 \pm 0.007)$ & $(178.6 \pm 0.1)$ \\ \hline
Res 2 & $400$ & $5.5379$ & $(10.40 \pm 1.75) \times 10^5$ & $(5955 \pm 4)$ & 2.496 & $(-32.788 \pm 0.005)$ & $(2.905 \pm 0.007)$ & $(130.1 \pm 0.2)$ \\ \hline
Res 3 & $500$ & $5.9991$ & $(10.66 \pm 1.65) \times 10^5$ & $(5850 \pm 1)$ & 2.232 & $(-25.354 \pm 0.006)$ & $(2.854 \pm 0.007)$ & $(98.6 \pm 0.1)$ \\ \hline
Res 4 & $600$ & $6.4067$ & $(4.16 \pm 0.64) \times 10^5$ & $(4544 \pm 1)$ & 2.038 & $(-21.259 \pm 0.006)$ & $(2.853 \pm 0.007)$ & $(82.4 \pm 0.1)$ \\ \hline
Res 5 & $700$ & $6.8399$ & $(3.83 \pm 0.31) \times 10^5$ & $(3421 \pm 1)$ & 1.887 & $(-19.179 \pm 0.004)$ & $(2.851 \pm 0.006)$ & $(70.1 \pm 0.2)$ \\ \hline
\end{tabular}
\caption{Summary table reporting the main parameters extracted for of all the grAl resonators with $w$ the resonator width, $f_0$ the resonant frequency, $Q_i$ the internal quality factor, $\langle n_{ph} \rangle$ the average intracavity photon number, $Q_c$ the coupling quality factor, $Z$ the characteristic impedance, $K$ the self-Kerr, and $B_{C,\parallel}$, $B_{C,\perp}$ the in-plane and out-of-plane critical magnetic field respectively.}
\label{table:grAl_resonators}
\end{table*}
\end{center}

Magnetic field sweeps were conducted using a standardized routine to ensure reliable and automated measurements. Initially, the resonant frequencies of each resonator at zero field were recorded as reference points. Each field sweep followed this procedure:
(1) A spectroscopy scan was performed at zero field with an input power corresponding to approximately 100 intracavity photons.
(2) The magnetic field was ramped up at a controlled rate of 100~mT/min to the next measurement point.
(3) After reaching the target field, a waiting time of 2 minutes was implemented to allow thermal stabilization, as the magnet ramping introduces heat at the 4~K stage.
(4) The next spectroscopy scan was performed.

For each resonance, a fast scan was initially conducted over a 80~MHz window below the previous resonant frequency to locate the new resonance. A detailed spectroscopy scan was then centered at the new resonant frequency, ensuring accurate tracking during the field sweep.

At predefined field intervals in the magnetic field sweep, power scans were also performed for each resonator. These scans provide insights into the evolution of the internal quality factor of the resonator under varying average intracavity photon populations.

Upon completing the field sweep, the magnet was ramped down gradually, ensuring the zero-field crossing was smooth to minimize hysteresis effects. However, residual vortices can remain trapped in the film, causing small shifts in resonant frequencies and variations in $Q_i$. To fully reset the device and eliminate residual vortices, the MXC stage was briefly heated above the critical temperature of the film before cooling back down to base temperature.

\subsection{\label{app:bfield_misalignment}Magnet misalignment estimation} 

Due to possible misalignment between the sample mount and/or the magnet within the cryostat, the applied magnetic field may have an unintended out-of-plane component. Accurately estimating this misalignment is essential, as even a small angular deviation can introduce significant out-of-plane fields when ramping up the in-plane field.

Using Eq.~\eqref{eq:freq_shift_bfield_inplane}, the misalignment angle $\theta_B$ can be extracted by analyzing how the resonator resonant frequency shifts $\Delta \omega / \omega_0$ depend on resonator width $w$. Since all resonators share the same device thickness, variations in $\Delta \omega / \omega_0$ with width reveal the effect of misalignment. If the applied in-plane field is sufficiently strong, this effect becomes measurable, as observed for NbN devices in Fig.~\ref{fig:nbn_bfield}\textbf{(a)}.

It is convenient to rewrite Eq.~\eqref{eq:freq_shift_bfield_inplane} as:
\begin{equation}
    \Delta \omega / \omega_0 = -\frac{\pi}{48} \frac{e^2t^2}{\hbar k_B T_C} D_p B_{\parallel}^2,
    \label{eq:freq_shift_bfield}
\end{equation}
where $D_p = D(1+\theta_B^2 \frac{w^2}{t^2})$. 
A linear fit of the extracted $D_p$ for each resonator with respect to $(w/t)^2$ provides the misalignment angle $\theta_B = 1.08^\circ$ [see Fig.~\ref{fig:nbn_misalignement}]. For grAl devices, the applied in-plane magnetic field is not strong enough due to the lower $B_{C,\parallel}$, making this effect less pronounced. Consequently, the misalignment for grAl devices is assumed to be the same as for the NbN case.

\section{\label{app:supp}Supplementary information}

Detailed information about all the resonator parameters are summarized in Tab.~\ref{table:NbN_resonators} and Tab.~\ref{table:grAl_resonators} for the NbN and grAl devices, respectively.

\subsection{\label{app:spec-in_bfield}Spectroscopy with magnetic field}

To provide better visualization of the $Q_i$ degradation under out-of-plane magnetic fields $B_\perp$, the resonance magnitude $|S_{21}|$ for the 300~nm wide NbN and grAl resonators taken for different values of $B_\perp$ are shown on the same plot in Figs.~\ref{fig:bfield_spectros}\textbf{(a)} and \ref{fig:bfield_spectros}\textbf{(b)}, respectively. 
For NbN, the resonance depth exhibits an exponential and continuous reduction as the magnetic field increases, indicating a gradual decline in $Q_i$. In contrast, the grAl resonators display a more stable resonance depth with occasional abrupt variations. Since the resonance depth is directly correlated with $Q_i$, these observations highlight the distinct responses of the two materials to out-of-plane magnetic fields, despite having identical device designs and geometries.

\subsection{\label{app:bfield_perp_offset}GrAl $Q_i$ evolution in out-of-plane magnetic field}

For improving visualization, the data in Fig.~\ref{fig:gral_bfield}\textbf{(d)} is replotted in Fig.~\ref{fig:gral_bfield_offset} with a y-axis offset for each dataset. This adjusted perspective highlights the maximum out-of-plane magnetic field value, $B_{\perp, \text{max}}$ (previously referred to as the "cutoff" field), beyond which $Q_i$ rapidly decreases. The dependence of $B_{\perp, \text{max}}$ on resonator width $w$ is shown in the inset, revealing that narrower resonators exhibit greater resilience to out-of-plane magnetic fields, as expected.

\subsection{\label{app:bfield_qc}External quality factor evolution with B field}

For completeness, alongside the internal quality factor $Q_i$ evolution under in-plane and out-of-plane magnetic fields (shown in Fig.~\ref{fig:nbn_bfield}\textbf{(c-d)} and Fig.~\ref{fig:gral_bfield}\textbf{(c-d)} for NbN and grAl resonators, respectively), the evolution of the external quality factor ($Q_c$) is presented in Fig.~\ref{fig:bfield_qc}\textbf{(a-d)}.
Overall, $Q_c$ remains relatively stable throughout the magnetic field sweeps, though increased fluctuations are observed near the end of the sweeps due to lower signal to noise ratio in this region. Notably, in NbN resonators, the $Q_c$ peak aligns with the ESR dip, further supporting the hypothesis of coupling to paramagnetic impurities. Additionally, variations in $Q_c$ may provide insights into potential coupling to standing waves in the feedline, which could influence $Q_i$ measurements.

\bibliography{Bibliography.bib,Bibliography_Resonators.bib,Bibliography_QuantumDots.bib,Bibliography_Qubits.bib,Bibliography_JPA,Bibliography_KIPA,Zotero.bib}

\end{document}